\documentclass[11pt,a4paper]{article}
\pdfoutput=1

\usepackage{jheppub}

\usepackage{amsmath,amssymb}
\usepackage{graphicx}
\usepackage{mciteplus}

\arxivnumber{1203.6636}

\title{Discrete family symmetry, Higgs mediators and  $\theta_{13}$}
\author[a]{Ivo de Medeiros Varzielas}
\emailAdd{ivo.de@udo.edu}
\affiliation[a]{Fakult\"{a}t f\"{u}r Physik, Technische Universit\"{a}t Dortmund
D-44221 Dortmund, Germany}
\author[b]{Graham G. Ross}
\emailAdd{g.ross1@physics.ox.ac.uk}
\affiliation[b]{Rudolf Peierls Centre for Theoretical Physics,
University of Oxford, 1 Keble Road, Oxford, OX1 3NP, U.K.}


\abstract{
We present a new (supersymmetric) framework for obtaining an excellent description of quark, charged lepton and neutrino masses and mixings from a $\Delta(6n^2)$ family symmetry with multiplet assignments consistent with an underlying $SO(10)$ Grand Unification. It employs a Higgs mediator sector in place of the usual Froggatt-Nielsen messengers, with quark and lepton messengers, and provides significant improvements over existing models of this type having unsuppressed Yukawa couplings to the third generation and a simplified vacuum alignment mechanism.  The neutrino mass differences are naturally less hierarchical than those of the quarks and charged leptons. Similarly the lepton mixing angles are much larger than those in the quark sector and have an approximate tri-bi-maximal (TB) mixing form for $\theta_{12}$ and $\theta_{23}$. However the mixing angle $\theta_{13}$ is naturally much larger than in pure TB  mixing and can be consistent with the value found in recent experiments. The magnitude of $\theta_{13}$ is correlated with a the predicted deviation of $\theta_{23}$ from bi-maximal mixing. The model has light familon fields that can significantly modify the associated SUSY phenomenology.
}

\begin{document}

\maketitle

\section{Introduction \label{Intro}}

Family symmetries (FSs) provide a promising framework for generating viable quark, charged lepton and neutrino masses and mixing. Indeed implementing the see-saw mechanism for neutrino masses in a SUSY model with non-Abelian discrete symmetries has been shown to lead quite naturally to near TB neutrino mixing while preserving an hierarchical structure for quark mixing angles and masses. \cite{Ivo1,Ivo3,Federica} belong to a specific framework of particular interest for unified models, other types of models are reviewed e.g. in \cite{Altarelli:2010gt} and present in the references therein. In most models, the fermion mass structure is generated through spontaneous breaking of the FS via the vacuum expectation values (vevs) of a limited number of familon fields. Fermions are massless in the FS limit and symmetry breaking is communicated to the fermion fields via the Froggatt-Nielsen mechanism with massive vectorlike pairs of fermion messenger fields.

Here we explore an alternative mediator sector composed of massive Higgs fields. In contrast to the fermion messenger case in \cite{Ivo1,Ivo3,Federica}, the third generation of fermions get a mass at renormalisable order. This results in a simplified familon sector and allows more naturally for the large top quark mass. We illustrate the alternative mechanism in a SUSY model with a discrete $\Delta(6 n^2)$ symmetry \cite{Escobar:2008vc, *Ishimori:2010au, *Escobar:2011mq} with quark and lepton fields assigned to FS multiplets in a manner consistent with an underlying $SO(10)$ GUT. We show that a specific $\Delta(6 n^2)$  multiplet assignment, supplemented by a simple $Z_{5} \times Z_4^R$ symmetry, leads to a viable structure for {\it all} quark and lepton masses and mixing. The significant difference between the quark and neutrino mixing angles is due to the fact that the latter are generated by the see-saw mechanism in which the third generation right-handed Majorana mass is very large, suppressing the would-be dominant third generation masses. The resulting neutrino mixing angles have an approximate TB mixing form. However the neutrino contribution to $\theta_{13}$, which is zero in TB mixing, has a significant correction that can lead to a naturally large value for $\theta_{13}$ consistent with the central value of the recent experimental measurements. Interestingly the value of $\theta_{13}$  is correlated with a significant departure from bi-maximal mixing of $\theta_{23}$. 

The paper is organised as follows. In Section \ref{sec:AFN} we introduce the multiplet structure and couplings of the model. Assuming a definite structure for vacuum alignment (familon vevs), we show that the model leads to consistent quark and lepton masses and mixing angles with approximate TB mixing in the lepton sector.  Section \ref{sec:t13} presents a discussion of how a significant  $\theta_{13}$ arises naturally and determines its correlation  with the deviation of $\theta_{23}$ from its bi-maximal value, while preserving the tri-maximal mixing structure associated with $\theta_{12}$. In Section \ref{sec:vevs} we complete the model, showing that the scalar potential consistent with the symmetries of the model provides the needed vacuum alignment. Section \ref{sec:R} discusses the phenomenological implications of the R-symmetry employed, particularly its implication for nucleon decay and other baryon- and lepton-number violating processes.  Finally Section \ref{sec:summary} presents a summary and our conclusions.

\section{A $\Delta(6 n^2)$ family symmetry model with Higgs mediators \label{sec:AFN}}

The model is based on $N=1$ supersymmetry with the Standard Model gauge group. In order to achieve phenomenologically viable structures for the fermions with near TB mixing for the leptons, we aim for fermion structures similar to those in \cite{Ivo1, Ivo3}, requiring the model to be consistent with an underlying $SO(10)$ unification with the left- and right-handed (L and R) sectors transforming in the same way under the FS.
This has the advantage that it is easy to obtain the phenomenologically successful $(1,1)$ texture zero relation relating the Cabibbo angle to the light quark masses \cite{Gatto, *Weinberg:1977hb, *Wilczek:1977uh, *Fritzsch:1977za}.
For notational simplicity it is convenient to label the representations by their transformation properties under the $SU(4)\times SU(2)_{L}\times SU(2)_{R}$ subgroup of $SO(10)$, although here we are only concerned with the $SU(3)\times SU(2)\times U(1)$ Minimal Supersymmetric Standard Model (MSSM) structure: extending this to $SU(4)$ would require a discussion of the $SU(4)$ symmetry breaking sector which is beyond the scope of this paper. The chiral supermultiplet structure is given in Table \ref{ta:ZN} where, in addition to the $\Delta(6 n^2)$ FS \cite{Escobar:2008vc, *Ishimori:2010au, *Escobar:2011mq}, we have allowed for a $Z_{5} \times Z_4^R$ discrete symmetry that restricts the allowed couplings.

\begin{table}
\begin{center}
\begin{tabular}{|c|ccc|ccc|}
\hline
Field & $SU(4)$ & $SU(2)_{L}$ & $SU(2)_{R}$ & $\Delta(6 n^2)$ & $Z_{5}$ & $Z_4^R$\\ \hline
$\Psi$ & $4$ & $2$ & $1$ & $3_{1_l}$ & $0$ & $1$\\
$\Psi^c$ & $\bar{4}$ & $1$ & $2$ & $3_{1_l}$ & $0$ & $1$\\ 
$\theta$ & $10$ & $1$ & $3$ & $1$ & $1$ & $0$\\ \hline
$H$ & $1$ & $2$ & $2$ & $1$ & $1$ & $0$ \\
$X$ & $1$ & $2$ & $2$ & $3_{1_{-2l}}$ & $0$ & $0$\\
$\bar{X}$ & $1$ & $2$ & $2$ & $3_{1_{2l}}$ & $0$ & $2$\\
$Y$ & $1$ & $2$ & $2$ & $3_{1_{-2l}}+3_{1_{l}}$ & $2$ & $0$\\
$\bar{Y}$ & $1$ & $2$ & $2$ & $3_{1_{2l}}+3_{1_{-l}}$ & $3$ & $2$\\
$Z$ & $1$ & $2$ & $2$ & $3_{2_{l}}$ & $0$ & $0$\\
$\bar{Z}$ & $1$ & $2$ & $2$ & $3_{1_{-l}}$ & $0$ & $2$ \\
$\Sigma$ & $15$ & $1$ & $1$ & $1$ & $3$ & $0$ \\ \hline
$\phi$ & $1$ & $1$ & $1$ & $ 1' $ & $0$ & $0$\\ 
$\phi_{1}$ & $1$ & $1$ & $1$ & $3_{1_l}$ & $1$ & $0$\\ 
$\bar\phi_{3}$ & $1$ & $1$ & $1$ & $3_{1_{-l}}$ & $2$ & $0$\\ 
$\bar\phi_{23}$ & $1$ & $1$ & $1$ & $3_{1_{-l}}$ & $3$ & $0$\\ 
$\bar\phi_{123}$ & $1$ & $1$ & $1$ & $3_{2_{-l}}$ & $1$ & $0$\\ 
\hline
\end{tabular}
\caption{Field and symmetry content of the model. \label{ta:ZN}}
\end{center}
\end{table}

Here, $\Psi_i$ contain the L fermions and $\Psi^c_j$ the conjugates of the R fermions, the quarks and the leptons of a single family being assigned to the $4$ of $SU(4)$. The Higgs sector comprises the  vectorlike superfields $X_{i},\; \bar{X}^{i},\;Y^{(a,b)}_{i},\; \bar{Y}^{(a,b),i}$ with bare masses of $O(M)$ and $Z_{i},\; \bar{Z}^{i}$ with a similar mass acquired through the vev of $\phi$, together with the MSSM-like Higgs fields $H$.
The group $\Delta(6n^2)$ is a subgroup of $SU(3)$, isomorphic to $(Z_n \times Z_n) \rtimes S_3$ where $\rtimes$ denotes the semi-direct product. Our notation identifies through the subscript $l$ the triplet
representations of $\Delta(6n^2)$, $3_{1_l}$ and $3_{2_l}$ ($3_{a_l}^\star \equiv 3_{a_{-l}}$). The constraint $n > 3$ follows because it is necessary for there to be three
independent representations associated with the product $3_{1_l} \times 3_{1_l} = 3_{1_{2l}}+3_{1_{-l}}+3_{2_{-l}}$ (for $n \leq 3$ the
representations $2l$ and $-l$ are equivalent). We have chosen $\Delta(6n^2)$ rather than $\Delta(3n^2)$
because only the former has an antisymmetric $3_{1_{-l}}$ coupling involving the product $3_{1_l} \times 3_{2_l}$, in the sense that the resulting representations are $3_{2_{2l}}= (a_1 b_1, a_2 b_2, a_3 b_3)$,  $3_{2_{-l}}=(a_2 b_3 + a_3 b_2, a_3 b_1 + a_1 b_3, a_1 b_2 + a_2 b_1)$ and $3_{1_{-l}}=(a_2 b_3 - a_3 b_2, a_3 b_1 - a_1 b_3, a_1 b_2 - a_2 b_1)$,  with $3_{1_l} = (a_1, a_2, a_3)$, $3_{2_l} = (b_1, b_2, b_3)$.
C.f. $3_{i_l} \times 3_{i_l}$, the same products are in the $3_{1_{2l}}$, $3_{1_{-l}}$ and $3_{2_{-l}}$ representations respectively, when $3_{i_l} = (a_1, a_2, a_3)$ and $3_{i_l} = (b_1, b_2, b_3)$  \cite{Escobar:2008vc, *Ishimori:2010au, *Escobar:2011mq}. 
The superfield $\Sigma$ is a Georgi-Jarlskog field that we assume descends from a field transforming as a $45$ dimensional representation of $SO(10)$ so that $H \Sigma$ is an effective 120, as discussed in \cite{Liliana}. It decouples from the right handed neutrinos and couples three times more strongly to the conjugate R leptons than the conjugate R down quarks.  

As shown in Section \ref{sec:vevs} the familons (GUT singlet fields), $\bar\phi_A$, acquire vevs breaking the FS. These vevs are aligned by the underlying non-Abelian structure giving  $<\bar\phi_{3}>\propto(0,0,1),\;<\bar\phi_{23}>\propto(0,-1,1)$ and $<\bar\phi_{123}>\propto(1,1,1)$ and it is this alignment that generates near TB mixing in the lepton sector. The constants of proportionality depend on the soft supersymmetry breaking terms and the $O(1)$ constants that are not determined by the symmetry. In what follows we shall just fit these constants to give acceptable masses for the quarks, The constraints on these constants are given in eq.(\ref{eq:eps2}) and eq.(\ref{eq:eps3}).
Note that we expect the familons break the discrete symmetries at a high scale. For this reason we do not discuss the constraints of discrete anomaly cancellation because they can be cancelled by additional massive states that do not affect the low energy phenomenology.

On FS breaking the Higgs fields, $H$, mix with the mediators. Due to the underlying chirality of the system (see below) there is one pair of Higgs doublets that remain light. These acquire vevs and through the Yukawa couplings give rise to the desired pattern of fermion masses. To see how this comes about we consider the superpotential that is allowed by the symmetries of the model ($SU(4)$ indices  are suppressed for clarity):
\begin{align}
P_S & = M \bar{X} X + M \bar{Y} Y+ \phi \bar{Z} Z+\bar\phi_3^{i}\bar\phi_3^{i}H \bar{X}^{i}/M_{X}^{a}\notag \\
&+a\bar\phi_{23}^{i}\bar\phi_{23}^{i}H \bar{Y}^{(a),i}/M_{X}^{b} +b[\bar\phi_{23}\bar\phi_{23}H \bar{Y}^{(b)}]_{+}/M_{X}^{b}
+[\bar\phi_{23} \bar\phi_{123} H \bar{Z}]_{-}/M_{X}^{c} \,,
\label{P_S}
\end{align}
where $\bar{Y}^{(a,b)}$ are the components of the field $\bar{Y}$ transforming as $3_{1_{2l}}$ and $3_{1_{-l}}$ respectively, with $[...]_{+,-}$ representing the symmetric or anti-symmetric contractions that make the terms $\Delta(6n^2)$ invariant, and where the MSSM $\mu$ term, that is of the same order as the soft supersymmetry breaking terms\footnote{The appearance of a $\mu$ term of the correct magnitude after supersymmetry breaking is guaranteed by a discrete R-symmetry as discussed in \cite{ZNR1, *ZNR2}.}, is relatively small and can, to a good approximation, be neglected here. Here the familon fields refer to their vevs. The masses  $M_{X}^{a}$, which can be the Planck, string or Grand Unified scale, are the masses of the messengers responsible for generating the non-renormalisable terms and we have absorbed the $O(1)$ couplings associated with these operators in the masses. Strictly we should include the $M_{X}$ scale messengers in our spectrum but we have in mind that these are all very heavy and belong to the fully unified theory while the fields $X,\;\bar{X},\;Y,\;\bar{Y},\;Z,\;\bar{Z}$ are the only ones lighter than $M_{X}$. Their lightness can readily be explained if their masses are initially forbidden by a chiral symmetry and arise through spontaneous breaking of the symmetry. Here we do not construct the full UV complete theory which requires the full $SO(10)$ invariant construction. For clarity we have written all their masses as $M$ because for $M<<M_{X}$ the differences between the masses are unimportant. The mass of the $Z,\;\bar{Z}$ pair comes from the vev of the $\Delta(6n^2)$ non-trivial singlet field $\phi$, as $3_{1_{l}} \times 1' = 3_{2_{l}}$, with the invariants arising from combining a triplet with its conjugate triplet e.g. $3_{i_{l}} \times 3_{i_{-l}} \rightarrow 1$ \cite{Escobar:2008vc, *Ishimori:2010au, *Escobar:2011mq}.

Note that these couplings respect a chiral symmetry under which $X$, $Y$, $Z$ and $H$ are even and $\bar{X}$, $\bar{Y}$ and $\bar{Z}$ are odd. As a result one combination of electroweak doublet fields, $H_l$, will be left light. If $M<<(\bar\phi_{3}^3)^{2}/M_{X}$ these are given by  
\begin{align}
H_{l} & \approx X_{3}+ \left( \frac{\bar\phi_{23}^3}{\bar\phi_{3}^3} \right)^{2}\frac{M_{X}^{a}}{M_{X}^{b}} \left( a( Y^{(a)}_{2}+Y^{(a)}_{3})-2b Y^{(b)}_{1} \right) \notag \\ 
&+ \left( \frac{\bar\phi_{23}^3 \bar\phi_{123}^3}{(\bar\phi_{3}^{3})^2} \right)\frac{M_{X}^{a}}{M_{X}^{c}}(2 Z_{1}-Z_{2}-Z_{3}) - H \frac{M\;M_{X}^{a}}{(\bar\phi_{3}^3)^{2}} \,,
\label{light}
\end{align}
After supersymmetry breaking the light Higgs fields acquire vevs in the usual manner as in the MSSM and fermion masses are generated. The underlying Yukawa couplings allowed by the symmetries of the model are
\begin{equation}
P_Y = X_{i} \Psi_i \Psi^c_i + \left( a' Y^{(a)}_{i} \Psi_{i}\Psi^c_{i} + b' [Y^{(b)} \Psi \Psi^c]_{+} \right) \Sigma/M_{X}+ [Z \Psi \Psi^c]_{-}
\end{equation}
where we have implicitly introduced a single pair of Froggatt-Nielsen matter messengers \cite{Froggatt:1978nt} which enables the coupling of $\Sigma$ to $\Psi^c$.


The Yukawa couplings and the light Higgs combinations lead to fermion Dirac mass matrices of the form
\begin{equation}
M_{f}\sim\left( \begin{matrix}
   0 & -\epsilon_{f}^{3} & \epsilon_{f}^{3}  \\
   \epsilon_{f}^{3} & a a' a^f \epsilon_{f}^{2} &-2b b' a^f \epsilon_{f}^{2} + 2 \epsilon_{f}^{3}  \\
   -\epsilon_{f}^{3} & -2b b' a^f\epsilon_{f}^{2} -2\epsilon_{f}^{3} & 1  \\
\end{matrix} \right) \,.
\label{dir}
\end{equation}
To fit the quark masses we have assumed the vevs and messenger masses then satisfy the relations
\begin{equation}
\epsilon_{f}^{2}=\left( \frac{\bar\phi_{23}^3}{\bar\phi_{3}^3} \right)^{2} \frac{M_{X}^{a}}{M_{X}^{b}}
\label{eq:eps2}
\end{equation}
\begin{equation}
\epsilon_{f}^{3} = \left(\frac{\bar\phi_{23}^3 \bar\phi_{123}^3}{(\bar\phi_{3}^{3})^2} \right)\frac{M_{X}^{a}}{M_{X}^{c}} \,,
\label{eq:eps3}
\end{equation}
Note that although we have used $SU(4)\times SU(2)_{L}\times SU(2)_{R}$ group representations for the quarks and leptons we expect the group (particularly $SU(2)_R$) to be strongly broken to the Standard Model gauge group, so we allow for different expansion parameters in the up and down quark sectors as is needed to explain the different quark mass hierarchies in these sectors \cite{Ivo1, Ivo3}. However, it may be that the breaking of the underlying $SO(10)$ group does not strongly break the equality of the expansion parameters in the up and down sectors separately. Here we make this simplifying assumption, labelling the expansion parameter in eq.(\ref{eq:eps2}) and eq.(\ref{eq:eps3}) as  $\epsilon\sim 0.15$ for the down quark and charged lepton sector, and as $\epsilon'\sim 0.05$ for the up quark and neutrino sector (these values have been shown to give a good description of the masses in these sectors). We can readily relax the assumption that the expansion parameter in the neutrino Dirac mass matrix is the same as in the up quark mass matrix without changing the predictions for the neutrino oscillation parameters. To explain the difference between $\epsilon$ and $\epsilon'$ requires $\frac{M_{X}^{a}}{M_{X}^{b}}|_{d}\approx 10 \frac{M_{X}^{a}}{M_{X}^{b}}|_{u}$. We introduce the convenient mass scale $M_d$ which is useful when comparing magnitudes between terms. $M_d$ is simultaneously associated with the hierarchy in the down (and charged lepton) sector and with the $\phi_{23}$ vev:
\begin{equation}
\epsilon \equiv \phi_{23}/M_d \,.
\label{eq:eps}
\end{equation}

The remaining parameters in eq(\ref{dir}) are given by $a^f\propto <\Sigma>/M$ with $a^{\nu}\sim 0,\;a^{l}\sim3a^{d}\sim3a^{u}/2$ for the neutrino, charged lepton, down and up quark mass matrices respectively. Note that the $(3,3)$ entries are generated at tree level, unsuppressed by inverse powers of the mediator mass (this is to be compared to the models of  \cite{Ivo1, Ivo3, KingPSL1, KingPSL2, SU3SU3}, where the $(3,3)$ entry is suppressed by one or more powers of the mediator mass).

As discussed in \cite{RRRV, *Mario} these Dirac mass matrices give acceptable masses and mixings for the quarks and acceptable masses for the charged leptons. The $(1,1)$ texture zero enables the successful relation \cite{Gatto, *Weinberg:1977hb, *Wilczek:1977uh, *Fritzsch:1977za} between the light quark masses and the Cabibbo angle.

\begin{figure}
\begin{center}
 \includegraphics[width=12 cm,keepaspectratio=true]{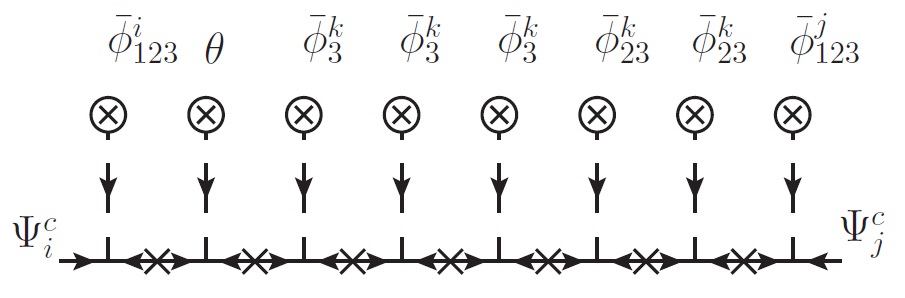}
\end{center}
\caption{An example of a subdominant Majorana diagram.}
\label{123_123}
\end{figure}

The light neutrinos get mass through the see-saw mechanism so we turn now to consider the structure of the Majorana masses of the right handed neutrinos. Since these violate lepton number it is necessary to introduce a source of lepton number violation and we adapt the scheme used in \cite{Ivo1, Ivo3}. We add the field $\theta$ which, c.f. Table \ref{ta:ZN}, has lepton number 2 and spontaneously breaks lepton number. It acquires a vev, close to the Planck scale, through radiative breaking. $\theta$ can be a single field or an effective field $\theta \propto\bar\theta \bar\theta$ where $\bar\theta$ is a $\bar{4}$ of $SU(4)$, and in this case the lepton number violation arises from the sneutrino vev of $\bar\theta$.
For $n=5$, the leading Majorana terms allowed by the symmetries are
\begin{align}
P_M & = \theta \Psi^c_{i} \Psi^c_{j} \left( \frac{\bar \phi_{3}^i\bar \phi_{3}^j}{M_{X}^2}
+ \frac{\bar \phi_{23}^i\bar \phi_{23}^j  \left[ \bar\phi_{3} \bar\phi_{3} \bar\phi_{23} \bar\phi_{23} \bar\phi_{23} \right]}{M_{X}^7}  \right) \notag \\
&+ \theta  \Psi^c_{i} \Psi^c_{j} \left( \frac{\bar\phi_{123}^i \bar\phi_{123}^j \left[ \bar\phi_{3} \bar\phi_{3} \bar\phi_{3} \bar\phi_{23} \bar\phi_{23} \right]}{M_{X}^7}+ \frac{\bar\phi_{23}^i \bar\phi_{123}^j \left[ \bar\phi_{3} \bar\phi_{3} \bar\phi_{3} \bar\phi_{23} \bar\phi_{123} \right]}{M_{X}^7} \right) \,,
\label{maj}
\end{align}
where $\left[...\right]$ represents the $\Delta(6 n^2)$ invariant given by $\sum_{i} \left [ \bar\phi_{A1}^{i} (...) \bar\phi_{An}^{i} \right]$.
We have not shown any of the allowed subdominant terms that only contribute to the third row or column, as they are not phenomenologically relevant due to the strong hierarchy between them and the first term.
In eq.(\ref{maj}) there are $O(1)$ coefficients associated with the couplings and messenger masses that are implicit. An example of a diagram generating one of the subdominant terms in eq.(\ref{maj}) is presented in Fig.(\ref{123_123}).
In an explicit completion the position of $\theta$ is not random as it depends on the gauge group representations of the available messengers.
In Fig.(\ref{123_123}), the position $\theta$ was chosen to better illustrate the contraction of the family index $i$.
To ensure that the model has a consistent UV completion, we carefully checked the symmetry content of the model and verified that the messengers required to enable the required Majorana terms do not conflict with e.g. the messenger content required for the correct vev alignment quartics.

Note that the Majorana masses are very strongly hierarchical. The third generation mass appears at $O \left( \frac{(\bar\phi_{3}^3)^{2}}{{M_X}^{2}}\theta \right)$ while the remaining masses are of $O \left(\frac{(\bar\phi_{3}^3)^2 (\bar\phi_{23}^3)^{5}}{M_{X}^{7}}\theta \right)$ and  $O \left(\frac{(\bar\phi_{3}^3)^3(\bar\phi_{23}^3)^{2}(\bar\phi_{123}^{3})^2}{M_{X}^{7}} \theta \right)$. As a result the dominant see-saw graphs involve the exchange of the two lighter states (the atmospheric and solar neutrino states) and, using the Dirac mass structure of eq(\ref{dir}), give light active neutrino masses that are related by 
\begin{equation}
\frac{m_{\odot}}{m_{@}}=\sqrt{\frac{\Delta m_{\odot}^2}{\Delta m_{@}^2}} \sim \frac{ (\bar\phi_{123}) ^2  (\bar\phi_{3}) }{ (\bar\phi_{23}) ^3}  \simeq O(\epsilon).
\label{eq:massratio}
\end{equation}
This naturally explains why the neutrino mass hierarchy of the two heaviest states is less than the corresponding hierarchy of the quark and charged leptons. The lightest neutrino is predicted to be very light due to the relatively large Majorana mass of the third generation RH neutrino\footnote{This strongly hierarchical Majorana mass hierarchy was christened ``Sequential Dominance'' in \cite{King:1999mb}.}.

What about the neutrino mixing angles? As discussed in \cite{Ivo1, Ivo3, Federica}, the see-saw mechanism with Majorana terms of the form of eq.(\ref{maj}), {\it without} the last term together with the Dirac masses  of eq.(\ref{dir}) generate the atmospheric and solar neutrino eigenstates $\nu_{@}\propto \sum_{i}\phi_{23}^i \nu_i$ and $\nu_{\odot}\propto \sum_{i}\phi_{123}^i \nu_i$ respectively. Since $\bar\phi_{23}\propto(0,-1,1)$ and $\bar\phi_{123}\propto(1,1,1)$ one immediately sees this corresponds to TB mixing.
However the last term perturbs the solution away from exact TB mixing and gives rise to a significant neutrino contribution to $\theta_{13}$.  We discuss its effect in detail in the next Section.

\section{Large $\theta_{13}$ \label{sec:t13}}

\begin{figure}
 \begin{center}
\includegraphics[width=5.7 cm,keepaspectratio=true]{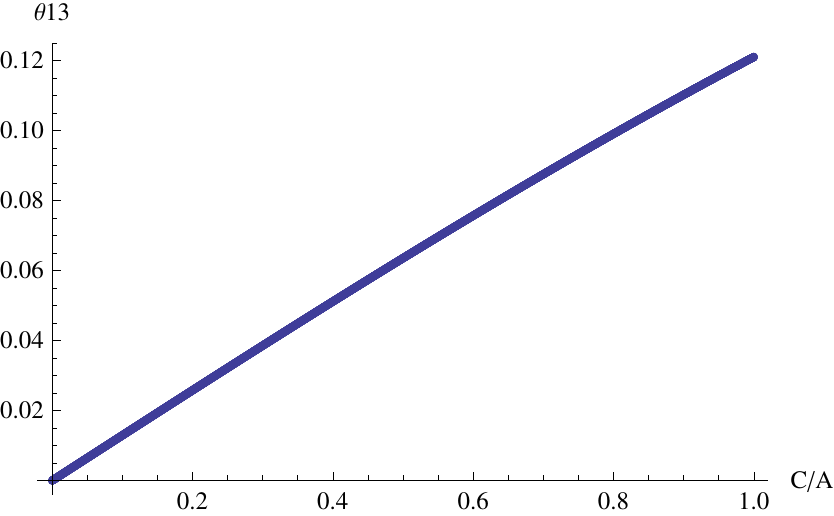}
 \includegraphics[width=5.7 cm,keepaspectratio=true]{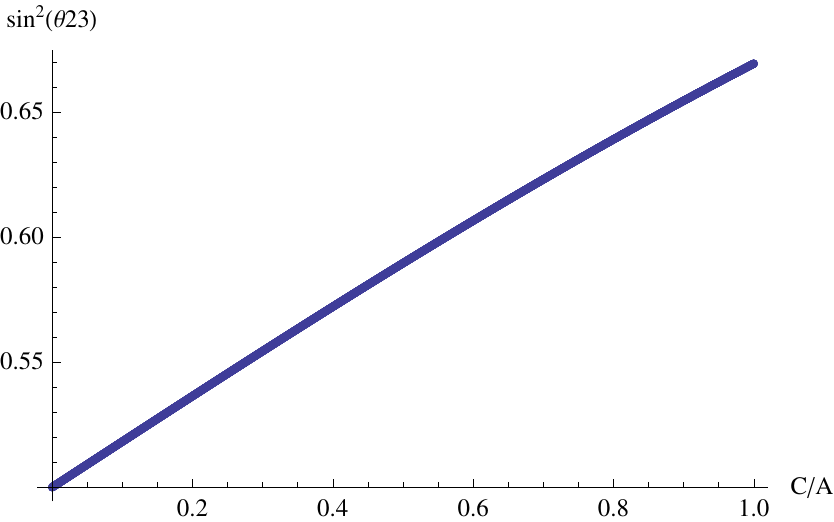}
 \includegraphics[width=5.7 cm,keepaspectratio=true]{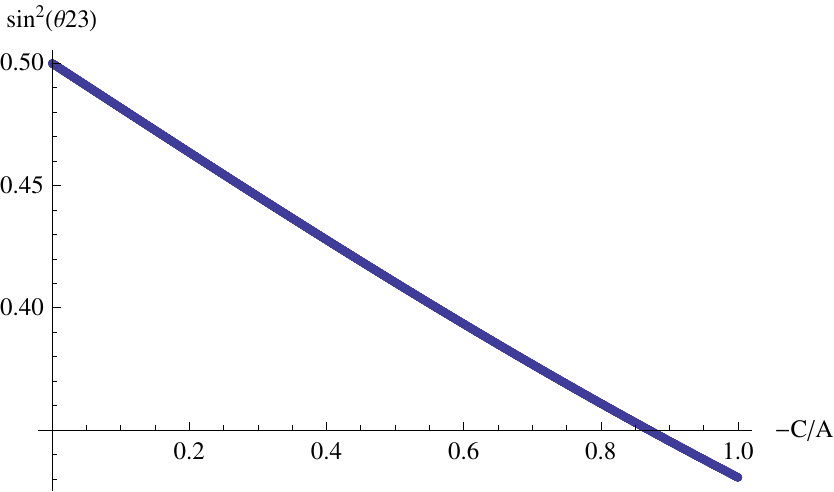}
\includegraphics[width=5.7 cm,keepaspectratio=true]{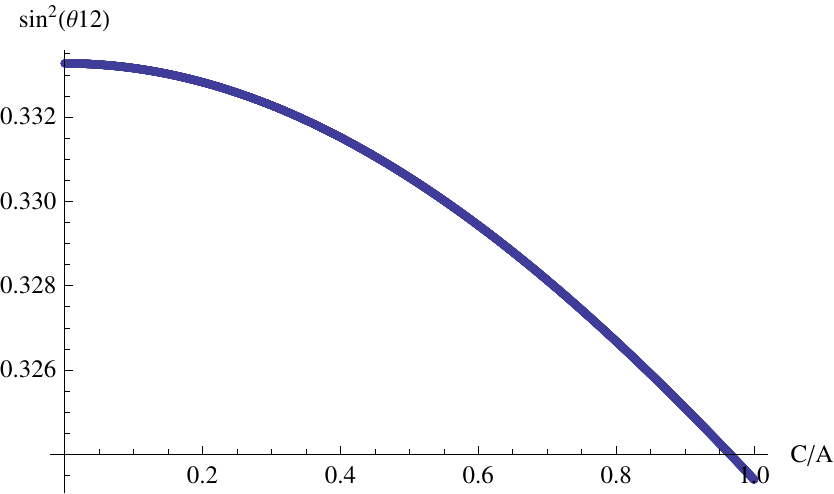}
 \end{center}
\caption{$\theta^\nu_{13}$, $\sin^2(\theta^\nu_{23})$ and $\sin^2(\theta^\nu_{12})$ plotted as a function of $C/A$.}
\label{plots}
\end{figure}

Recent observations indicate that $\theta_{13}$ is relatively large \cite{Abe:2011sj, *Fogli:2011qn, *Schwetz:2011zk, *Machado:2011ar}, \cite{DayaBay}. This is in some tension with the small contribution predicted in discrete FS models which have no contribution to the mixing angle from the neutrino sector and only a small contribution from the charged lepton sector given by:
\begin{equation}
\theta_{13} \simeq \theta_{C}/(3 \sqrt{2}) \simeq 0.05 .
\label{eq:t13GUT}
\end{equation}
Nonetheless, FSs remain extremely appealing and may naturally produce values of $\theta_{13}$ in agreement with the  central value of the recent Daya Bay measurement. Many different approaches have been considered \cite{He:2011kn, *Xing:2011at, *Ma:2011yi,  *Zheng:2011uz,  *Zhou:2011nu,  *Araki:2011wn,  *Haba:2011nv,  *Morisi:2011pm,  *Chao:2011sp,  *Zhang:2011aw,  *Dev:2011bd,  *Chu:2011jg,  *BhupalDev:2011gi,  *Toorop:2011jn,  *Antusch:2011qg,  *Rodejohann:2011uz,  *Ahn:2011if,  *King:2011zj,  *Marzocca:2011dh,  *Ge:2011qn,  *Kumar:2011vf,  *Bazzocchi:2011ax,  *Araki:2011qy,  *Antusch:2011ic,  *Fritzsch:2011qv,  *Rashed:2011zs,  *Ludl:2011vv,  *Verma:2011kz,  *Meloni:2011ac,  *Dev:2011hf,  *Deepthi:2011sk,  *Rashed:2011xe, *King:2011ab,  *Araki:2011zg,  *Gupta:2011ct,  *Ding:2012xx,  *Ishimori:2012gv,  *Dev:2012ns, *Bazzocchi:2012ve, *BhupalDev:2012nm, *Cooper:2012wf, *Siyeon:2012zu, *Wu:2012ri, *Branco:2012vs, *He:2012yt, *Meloni:2012ci, *Ahn:2012tv}.
Relevant to the present model, as discussed in \cite{SU3SU3}, in TB mixing models with type I seesaw one can introduce deviations through the vev structure itself (see e.g. \cite{King:2009qt}), the Yukawa structure (as in \cite{SU3SU3}) or through the Majorana structure. Here we consider the latter as a source of the deviation from pure TB mixing in the neutrino sector. In fact, as mentioned above, we have not discussed the last term in eq.(\ref{maj}) that breaks TB mixing. It is convenient to use the mass scale $M_d$ defined in eq.(\ref{eq:eps}), in order to express the relative magnitudes clearly in terms of $\epsilon$. Up to $O(1)$ coefficients this term has the same magnitude as the second to last term in eq.(\ref{maj}) and is suppressed by $O(M_d/M_{X})^5$ with respect to the dominant first term of eq.(\ref{maj}). 
The resultant Majorana mass matrix has the form
\begin{equation}
M_{R}\propto \left( \begin{matrix}
  B {{\epsilon }^{6}} & (B+C) {{\epsilon }^{6}} & .  \\
   (B+C) {{\epsilon }^{6}} & A {{\epsilon }^{5}} & .  \\
   . & . & (M_X/M_d)^5 \\ 
\end{matrix} \right) \,.
\label{MR}
\end{equation}
where the $(i,3)$ and $(3,i)$ entries have negligible effect due to the dominance of the $(3,3)$ entry and we have made explicit the $O(1)$ coefficients $A,B,C$ associated with the last three terms of eq.(\ref{maj}) -  these dimensionless coefficients $A,B,C$ appear at the same level so one does not expect their ratios to deviate from $O(1)$. Note that the $B,C$ terms have one additional $\epsilon$ compared to the $A$ term, which leads to eq.(\ref{eq:massratio}).  The largest Majorana mass does not affect neutrino oscillation phenomenology as it only enters in the determination of the lightest neutrino mass which is negligible.
Indeed, to a very good approximation, we can rotate the effective neutrino mass matrix $m_\nu$ with the TB mixing matrix to obtain:
\begin{equation}
m_{\nu}\propto \frac{1}{AB-C^2} \left( \begin{matrix}
  0 & 0 & 0  \\
  0 & 3 B & \sqrt{6} C  \\
  0 & \sqrt{6} C & 2 A/\epsilon \\ 
\end{matrix} \right) \,.
\label{mnuSD}
\end{equation}
This shows clearly that when $C=0$, $m_\nu$ is diagonalised by TB mixing with one vanishing eigenvalue, the other two being proportional respectively to $3/A$ (solar) and $2/ (B \epsilon)$ (atmospheric). When $C \neq 0$ an additional rotation of the $23$ sector is required to diagonalise $m_\nu$, and it is convenient to parametrize this additional rotation with the angle $\varphi$:
\begin{equation}
U_{\varphi}= \left( \begin{matrix}
  1 & 0 & 0  \\
  0 & \cos(\varphi) & \sin(\varphi) \\
  0 & -\sin(\varphi) & \cos(\varphi) \\ 
\end{matrix} \right) \,,
\label{extrarotation}
\end{equation}
where $\tan (2\varphi) = - 2 \sqrt{6} C/(3 B - 2 A/\epsilon) \approx \epsilon \sqrt{6}  C/A$.
The plots in Fig. \ref{plots} illustrate how varying the ratio $C/A$ controls the perturbation in the angles $\theta^\nu_{13}$, $\theta^\nu_{23}$ and $\theta^\nu_{12}$ that diagonalise the effective neutrino mass matrix \footnote{The coefficients $A,B$ are chosen to obtain the desired mass splitting between the atmospheric and solar neutrino masses.}.  As $C/A$ increases, the deviation in $\theta^\nu_{13}$ and $\theta^\nu_{23}$ from the TB mixing values increases linearly (at leading order in the small angle $\varphi$) and becomes significant. To obtain the central value of the Daya Bay experiment, $\theta_{13} \simeq 0.15$, assuming the contribution from diagonalising the charged leptons adds coherently, requires $C/A \simeq 0.8$ which changes $\sin^{2}\theta_{23}$ from the bi-maximal value of $0.5$ either to $0.6$ or to $0.4$ depending on the sign of $C/A$, consistent with present measurement. Note that $C$ and $A$ are unknown coefficients expected to be of $O(1)$ so a large value of $C/A$ is not unreasonable. As may be seen from Fig. \ref{plots} the mixing angle $\theta_{12}$ is barely changed from the tri-maximal mixing value, its deviation being at leading order quadratic in $C/A$ (consistent with doing an expansion in the small angle $\varphi$). Observation of the correlations between the leptonic mixing angles would provide a test of the model. In general we expect the $O(1)$ coefficients and the vevs to be complex which generate CP violation. The phases complicate the analysis of the mixing angles, but not the expectation that $\theta_{23}$ receives significant corrections and depending on the Dirac CP violating phase $\sin^2(\theta_{23})$ can be between what is obtained for the real values of $C/A$ of either sign. The model falls in a general class of models which perturb TB mixing and the implications for the Dirac CP violating phase will be discussed in more detail in a future work \cite{HallRoss}.

\section{Vacuum alignment \label{sec:vevs}}
In the absence of supersymmetry breaking the scalar potential has no dependence on the familon fields alone. Thus the vacuum alignment depends on supersymmetry breaking and the familon fields can acquire very large vevs along flat directions.  Even if the soft supersymmetry breaking mass squared terms for the familon fields are initially positive, they can be driven negative by radiative corrections triggering a vev for the field, therefore the familon fields may spontaneously break the family symmetry through radiative breaking \cite{IbanezRoss}. 
In this the familon mass squared, positive at the initial scale (which could be the Planck scale), is driven negative through its Yukawa coupling to the messenger fields. The scale at which this happens depends sensitively on the magnitude of these couplings.  Since there are no stabilising F-terms involving the familons the magnitude of the familon vevs is close to the scale at which their masses squared are driven negative which can be close to the initial scale. Similarly the relative magnitude of the various familon vevs is determined by these couplings and the initial value of the soft SUSY breaking parameters. Given our ignorance of the familon Yukawa couplings we treat the magnitude of the familon vevs as free parameters. At this stage the potential, only having the quadratic mass term in the familon fields, is invariant under a continuous $SU(3)$ family symmetry and so the relative alignment of a familon multiplet is not fixed. To determine this alignment we turn to a consideration of the leading, non-renormalisable terms that respect the discrete family symmetry but break the full $SU(3)$ symmetry.


Let us start with the field $\bar\phi_{3}$ that gets the largest vev. The soft SUSY breaking mass term in the potential, $m_{3}^{2}|\phi_{3}|^{2}$ is actually symmetric under the larger continuous $SU(3)$ FS so does not pick a particular direction for the vev when $m_{3}^{2}$ becomes negative. The leading allowed term that is invariant under $\Delta(6 n^2)$ but not $SU(3)$ is the D-term that arises in radiative order and is proportional to $m^{2}\sum_i \bar\phi_{3}^{\dagger i} \bar\phi_{3,i }\bar\phi_{3}^{\dagger i} \bar\phi_{3,i}/M_{X}^{2}$. Minimising this term requires $\bar\phi_{3}\propto(0,0,1)$  if the coefficient of this term is negative. For the case of $\phi_{1}$, the quartic $m^{2}\sum_i \bar\phi_{3}^{\dagger i} \bar\phi_{3,i }\phi_{1}^{\dagger i} \phi_{1,i}/M_{X}^{2}$ requires $\phi_{1}$ be orthogonal to $\phi_{3}$ if the coefficient of the term is positive and, if the pure $\phi_{1}$ quartic has a negative coefficient, it will align in the direction $\phi_{1}\propto(1,0,0)$ as required. 

The alignment of the familons $\bar\phi_{23}$ and $\bar\phi_{123}$ proceeds in a similar manner although with a different origin for the alignment terms. In this case there is a term allowed in the superpotential of the form $P(\bar\phi_{123})^{2}\bar\phi_{23}/M_{X}^{3}$ where $P$ is the superpotential with R-charge 2. On supersymmetry breaking $P$ acquires a vev such that $P/M_{X}^{2}=m$, generating the gravitino mass $m$, and giving the term  $m (\bar\phi_{123})^{2}  \bar\phi_{23}/M_{X}$ in the superpotential. The related F-terms give both a  pure quartic for $\bar\phi_{123}$  and a mixed quartic involving $\bar\phi_{23}$. As they are generated at tree level they dominate over radiatively induced quartic terms involving $\bar\phi_{3}$ and $\bar\phi_{23}$. Since their coefficients are positive they ensure that the $\bar\phi_{23},\;\bar\phi_{123}$ vevs are orthogonal and that $\bar\phi_{123}\propto (1,1,1)$. Finally, if the mixed quartic $m^{2}\sum_i \bar\phi_{1}^{\dagger i} \bar\phi_{1,i }\phi_{23}^{\dagger i} \phi_{23,i}/M_{X}^{2}$ has positive coefficient and dominates over the one involving $\bar\phi_{3}$, it requires $\bar{\phi}_{23}^1  = 0$ and hence $\bar\phi_{23}\propto (0,1,-1)$. 
This completes the alignment discussion. It represents a significant improvement over the alignment solutions presented in \cite{Ivo3, SU3SU3} as  the alignment  mechanism is relatively insensitive to the relative magnitude of the vevs.

\section{Phenomenological implications \label{sec:R}}

The supersymmetric model constructed here has a $Z_{4}^{R}$ symmetry that leads to some phenomenological differences compared to the MSSM. As discussed in \cite{ZNR1} supersymmetry breaking also breaks this symmetry to $Z_{2}^{R}$ which is equivalent to matter parity in the MSSM. As a result the LSP is stable and a dark matter candidate. However the symmetry has significant advantages over the MSSM in that the $\mu$ term is naturally of order the supersymmetry breaking scale in the visible sector and the dangerous dimension 5 operators that lead to nucleon decay are absent.   

The novel Higgs mediation introduced here requires the presence of some vectorlike electroweak doublet states with mass of $O(M)$. As we have discussed, familon vevs can be close to the Planck scale. In this case the constraint needed for the validity of eq(\ref{light}), $M<<(\bar\phi_{3}^3)^{2}/M_{X}$, can be satisfied for large values of $M$,  even for $M_{X}$ close to the Planck mass. As a result the new vectorlike electroweak doublet states may be quite close to the unification scale and may not destroy the success of gauge coupling unification. 

A feature of the alignment scheme discussed here in which the alignment proceeds via the supersymmetry breaking terms only is that the familons will be very light. In the following, unlike in previous sections, we explicitly distinguish the components of the familons from their respective vevs. All the scalar components of the familon fields are expected to acquire masses of order of the gravitino mass from the usual supersymmetry breaking D-terms. However the fermion components can be much lighter as they can only get mass from F-terms.
The fermion component masses come from the superpotential term $m\bar\phi_{123}^{2}\bar\phi_{23}$. To leading order in $\epsilon$ this gives Majorana masses $m \langle \phi_{23} \rangle \sim m\epsilon$ to two fermion states of $\bar\phi_{123}$. In addition, all components of $\bar\phi_{123}$ and $\bar\phi_{23}$ have Dirac masses $m \langle \bar\phi_{123} \rangle \sim m\epsilon^{2}$, giving one component of $\bar\phi_{23}$ and the remaining component of $\bar\phi_{123}$ mass $m\epsilon^{2}$.  The remaining fermion states of $\bar\phi_{23}$ get Majorana mass $(m\langle \bar\phi_{123} \rangle)^{2}/ m\langle \bar\phi_{23} \rangle\sim m\epsilon^{3}$ via a see-saw mechanism. 

The coupling of the familon states is highly suppressed. For example the Yukawa coupling of $\bar\phi_{3}$ to the third family of quarks is proportional to $\frac{\langle H_{l} \rangle}{\langle \bar\phi_{3} \rangle}\left ( \frac{MM_{X}}{\langle \bar\phi_{3} \rangle^{2}}\right )^{2}$. For $\langle \bar\phi_{3} \rangle$ close to the Planck mass these are Planck scale suppressed couplings so that the familons may be considered to be hidden sector fields and do not play a direct role in the low energy phenomenology of the model. However, given their low mass, it may be that a familon fermion state is the lightest supersymmetric state (LSP). In this case the ``visible sector'' LSP will ultimately decay into the familon LSP, despite its Planck suppressed coupling. This will have to be taken into account when determining the dark matter density in this model and, due to its Planck suppressed coupling, direct dark matter detection of such a familon LSP is not possible. Finally we note that SUSY phenomenology may be significantly affected by the existence of a familon LSP as there is no longer a requirement that the ``visible sector'' LSP should be neutral. Thus the missing energy signals that are usually associated with SUSY particle decay into a neutral LSP may be absent.

\section{Summary \label{sec:summary}}

In this paper we have explored the possibility that the spontaneous breaking of family symmetry is communicated to the quarks and leptons by electroweak doublet Higgs messenger fields that also transform under the family symmetry, rather than the usual Froggatt-Nielsen mechanism that involve messengers carrying quark and lepton quantum numbers. The new mechanism has the advantage that the third generation Yukawa couplings are unsuppressed by familon vevs, readily allowing for the large top quark mass. Starting with the MSSM with an extended $Z_{4}^{R}$ symmetry to control the $\mu$ term and nucleon decay, we showed that the observed fermion mass structure could be generated through a relatively simple $\Delta(6 n^2)\times Z_{5}$ family symmetry, with multiplet structure chosen to be consistent with an underlying $SO(10)$ GUT.  The mass structure follows from a particular choice of family symmetry breaking and we showed how the associated vacuum alignment of the familon vevs is naturally determined by the structure of the allowed radiative D-terms, a considerable simplification of the usual alignment mechanisms.

The model explains why the splitting between the atmospheric neutrinos is smaller than that of the quarks and leptons,  $\sqrt{\frac{\Delta m_{\odot}^2}{\Delta m_{@}^2}} \simeq \epsilon$ compared to $\frac{m_{s}}{m_{b}}\sim \epsilon^{2}$ where $\epsilon\sim 0.15$. It also predicts near tri-bi-maximal mixing in the neutrino sector for the solar and atmospheric mixing angles. However this structure is only approximate and allows for a sizeable neutrino contribution to $\theta_{13}$ with size (governed by undetermined $O(1)$ coefficients) that is correlated to a deviation of $\theta_{23}$ from its bi-maximal value. Allowing for a value of $\theta_{13}=0.15$, to agree with the central value found by the Daya Bay experiment, gives $\sin^2 \theta_{23} \simeq 0.4-0.6$. 

The vectorlike Higgs messenger fields are expected to  be very heavy, close to the Planck or unification scale, and decouple from low energy physics. However the familon fields are expected to be light, with masses of order the supersymmetry breaking scale in the visible sector. This follows because the simple alignment mechanism presented here involves only D-term radiatively induced operators that are proportional to the square of the supersymmetry breaking mass. The coupling of the familons to the MSSM states is strongly suppressed by a power of the familon vevs and so, despite being light, the familons belong to a decoupled sector and are not expected to play a direct role in low-energy experiments. After supersymmetry breaking, the model has the usual matter parity of the MSSM, implying that the LSP is stable and that superpartners can only be produced in pairs. However it is possible that a familon fermion is the LSP and, in this case, the low energy phenomenology can be significantly changed because the ``visible sector'' LSP will decay to the familon LSP. Thus it is possible for the ``visible sector'' LSP to be charged and the standard missing energy signals associated with the production of superpartners to be much reduced.

\acknowledgments

We would like to thank the BCTP at University of Bonn for hospitality during part of this work and Christoph Luhn for helpful discussions.
IdMV was supported by DFG grant PA 803/6-1 and partially supported through the project PTDC/FIS/098188/2008. GGR acknowledges support from the EU ITN grant UNILHC 237920, the ERC Advanced grant BSMOXFORD 228169 and the Leverhulme Trust.

\bibliographystyle{apsrevM}
\bibliography{refs}

\ifx\mcitethebibliography\mciteundefinedmacro
\PackageError{apsrevM.bst}{mciteplus.sty has not been loaded}
{This bibstyle requires the use of the mciteplus package.}\fi
\begin{mcitethebibliography}{76}
\expandafter\ifx\csname natexlab\endcsname\relax\def\natexlab#1{#1}\fi
\expandafter\ifx\csname bibnamefont\endcsname\relax
  \def\bibnamefont#1{#1}\fi
\expandafter\ifx\csname bibfnamefont\endcsname\relax
  \def\bibfnamefont#1{#1}\fi
\expandafter\ifx\csname citenamefont\endcsname\relax
  \def\citenamefont#1{#1}\fi
\expandafter\ifx\csname url\endcsname\relax
  \def\url#1{\texttt{#1}}\fi
\expandafter\ifx\csname urlprefix\endcsname\relax\def\urlprefix{URL }\fi
\providecommand{\bibinfo}[2]{#2}
\providecommand{\eprint}[2][]{\url{#2}}

\bibitem[{\citenamefont{de~Medeiros~Varzielas and Ross}(2006)}]{Ivo1}
\bibinfo{author}{\bibfnamefont{I.}~\bibnamefont{de~Medeiros~Varzielas}}
  \bibnamefont{and} \bibinfo{author}{\bibfnamefont{G.~G.} \bibnamefont{Ross}},
  \bibinfo{journal}{Nucl.Phys.} \textbf{\bibinfo{volume}{B733}},
  \bibinfo{pages}{31} (\bibinfo{year}{2006}), \eprint{hep-ph/0507176}\relax
\mciteBstWouldAddEndPuncttrue
\mciteSetBstMidEndSepPunct{\mcitedefaultmidpunct}
{\mcitedefaultendpunct}{\mcitedefaultseppunct}\relax
\EndOfBibitem
\bibitem[{\citenamefont{de~Medeiros~Varzielas
  et~al.}(2007)\citenamefont{de~Medeiros~Varzielas, King, and Ross}}]{Ivo3}
\bibinfo{author}{\bibfnamefont{I.}~\bibnamefont{de~Medeiros~Varzielas}},
  \bibinfo{author}{\bibfnamefont{S.~F.} \bibnamefont{King}}, \bibnamefont{and}
  \bibinfo{author}{\bibfnamefont{G.~G.} \bibnamefont{Ross}},
  \bibinfo{journal}{Phys. Lett.} \textbf{\bibinfo{volume}{B648}},
  \bibinfo{pages}{201} (\bibinfo{year}{2007}), \eprint{hep-ph/0607045}\relax
\mciteBstWouldAddEndPuncttrue
\mciteSetBstMidEndSepPunct{\mcitedefaultmidpunct}
{\mcitedefaultendpunct}{\mcitedefaultseppunct}\relax
\EndOfBibitem
\bibitem[{\citenamefont{Bazzocchi and de~Medeiros~Varzielas}(2009)}]{Federica}
\bibinfo{author}{\bibfnamefont{F.}~\bibnamefont{Bazzocchi}} \bibnamefont{and}
  \bibinfo{author}{\bibfnamefont{I.}~\bibnamefont{de~Medeiros~Varzielas}},
  \bibinfo{journal}{Phys.Rev.} \textbf{\bibinfo{volume}{D79}},
  \bibinfo{pages}{093001} (\bibinfo{year}{2009}), \eprint{0902.3250}\relax
\mciteBstWouldAddEndPuncttrue
\mciteSetBstMidEndSepPunct{\mcitedefaultmidpunct}
{\mcitedefaultendpunct}{\mcitedefaultseppunct}\relax
\EndOfBibitem
\bibitem[{\citenamefont{Altarelli and Feruglio}(2010)}]{Altarelli:2010gt}
\bibinfo{author}{\bibfnamefont{G.}~\bibnamefont{Altarelli}} \bibnamefont{and}
  \bibinfo{author}{\bibfnamefont{F.}~\bibnamefont{Feruglio}},
  \bibinfo{journal}{Rev.Mod.Phys.} \textbf{\bibinfo{volume}{82}},
  \bibinfo{pages}{2701} (\bibinfo{year}{2010}), \eprint{1002.0211}\relax
\mciteBstWouldAddEndPuncttrue
\mciteSetBstMidEndSepPunct{\mcitedefaultmidpunct}
{\mcitedefaultendpunct}{\mcitedefaultseppunct}\relax
\EndOfBibitem
\bibitem[{\citenamefont{Escobar and Luhn}(2009)}]{Escobar:2008vc}
\bibinfo{author}{\bibfnamefont{J.}~\bibnamefont{Escobar}} \bibnamefont{and}
  \bibinfo{author}{\bibfnamefont{C.}~\bibnamefont{Luhn}},
  \bibinfo{journal}{J.Math.Phys.} \textbf{\bibinfo{volume}{50}},
  \bibinfo{pages}{013524} (\bibinfo{year}{2009}), \eprint{0809.0639}\relax
\mciteBstWouldAddEndPuncttrue
\mciteSetBstMidEndSepPunct{\mcitedefaultmidpunct}
{\mcitedefaultendpunct}{\mcitedefaultseppunct}\relax
\EndOfBibitem
\bibitem[{\citenamefont{Ishimori et~al.}(2010)\citenamefont{Ishimori,
  Kobayashi, Ohki, Shimizu, Okada et~al.}}]{Ishimori:2010au}
\bibinfo{author}{\bibfnamefont{H.}~\bibnamefont{Ishimori}},
  \bibinfo{author}{\bibfnamefont{T.}~\bibnamefont{Kobayashi}},
  \bibinfo{author}{\bibfnamefont{H.}~\bibnamefont{Ohki}},
  \bibinfo{author}{\bibfnamefont{Y.}~\bibnamefont{Shimizu}},
  \bibinfo{author}{\bibfnamefont{H.}~\bibnamefont{Okada}},
  \bibnamefont{et~al.}, \bibinfo{journal}{Prog.Theor.Phys.Suppl.}
  \textbf{\bibinfo{volume}{183}}, \bibinfo{pages}{1} (\bibinfo{year}{2010}),
  \eprint{1003.3552}\relax
\mciteBstWouldAddEndPuncttrue
\mciteSetBstMidEndSepPunct{\mcitedefaultmidpunct}
{\mcitedefaultendpunct}{\mcitedefaultseppunct}\relax
\EndOfBibitem
\bibitem[{\citenamefont{Escobar}(2011)}]{Escobar:2011mq}
\bibinfo{author}{\bibfnamefont{J.}~\bibnamefont{Escobar}},
  \bibinfo{journal}{Phys.Rev.} \textbf{\bibinfo{volume}{D84}},
  \bibinfo{pages}{073009} (\bibinfo{year}{2011}), \eprint{1102.1649}\relax
\mciteBstWouldAddEndPuncttrue
\mciteSetBstMidEndSepPunct{\mcitedefaultmidpunct}
{\mcitedefaultendpunct}{\mcitedefaultseppunct}\relax
\EndOfBibitem
\bibitem[{\citenamefont{Gatto et~al.}(1968)\citenamefont{Gatto, Sartori, and
  Tonin}}]{Gatto}
\bibinfo{author}{\bibfnamefont{R.}~\bibnamefont{Gatto}},
  \bibinfo{author}{\bibfnamefont{G.}~\bibnamefont{Sartori}}, \bibnamefont{and}
  \bibinfo{author}{\bibfnamefont{M.}~\bibnamefont{Tonin}},
  \bibinfo{journal}{Phys.Lett.} \textbf{\bibinfo{volume}{B28}},
  \bibinfo{pages}{128} (\bibinfo{year}{1968})\relax
\mciteBstWouldAddEndPuncttrue
\mciteSetBstMidEndSepPunct{\mcitedefaultmidpunct}
{\mcitedefaultendpunct}{\mcitedefaultseppunct}\relax
\EndOfBibitem
\bibitem[{\citenamefont{Weinberg}(1977)}]{Weinberg:1977hb}
\bibinfo{author}{\bibfnamefont{S.}~\bibnamefont{Weinberg}},
  \bibinfo{journal}{Trans.New York Acad.Sci.} \textbf{\bibinfo{volume}{38}},
  \bibinfo{pages}{185} (\bibinfo{year}{1977})\relax
\mciteBstWouldAddEndPuncttrue
\mciteSetBstMidEndSepPunct{\mcitedefaultmidpunct}
{\mcitedefaultendpunct}{\mcitedefaultseppunct}\relax
\EndOfBibitem
\bibitem[{\citenamefont{Wilczek and Zee}(1977)}]{Wilczek:1977uh}
\bibinfo{author}{\bibfnamefont{F.}~\bibnamefont{Wilczek}} \bibnamefont{and}
  \bibinfo{author}{\bibfnamefont{A.}~\bibnamefont{Zee}},
  \bibinfo{journal}{Phys.Lett.} \textbf{\bibinfo{volume}{B70}},
  \bibinfo{pages}{418} (\bibinfo{year}{1977})\relax
\mciteBstWouldAddEndPuncttrue
\mciteSetBstMidEndSepPunct{\mcitedefaultmidpunct}
{\mcitedefaultendpunct}{\mcitedefaultseppunct}\relax
\EndOfBibitem
\bibitem[{\citenamefont{Fritzsch}(1977)}]{Fritzsch:1977za}
\bibinfo{author}{\bibfnamefont{H.}~\bibnamefont{Fritzsch}},
  \bibinfo{journal}{Phys.Lett.} \textbf{\bibinfo{volume}{B70}},
  \bibinfo{pages}{436} (\bibinfo{year}{1977})\relax
\mciteBstWouldAddEndPuncttrue
\mciteSetBstMidEndSepPunct{\mcitedefaultmidpunct}
{\mcitedefaultendpunct}{\mcitedefaultseppunct}\relax
\EndOfBibitem
\bibitem[{\citenamefont{Ross and Velasco-Sevilla}(2003)}]{Liliana}
\bibinfo{author}{\bibfnamefont{G.~G.} \bibnamefont{Ross}} \bibnamefont{and}
  \bibinfo{author}{\bibfnamefont{L.}~\bibnamefont{Velasco-Sevilla}},
  \bibinfo{journal}{Nucl.Phys.} \textbf{\bibinfo{volume}{B653}},
  \bibinfo{pages}{3} (\bibinfo{year}{2003}), \eprint{hep-ph/0208218}\relax
\mciteBstWouldAddEndPuncttrue
\mciteSetBstMidEndSepPunct{\mcitedefaultmidpunct}
{\mcitedefaultendpunct}{\mcitedefaultseppunct}\relax
\EndOfBibitem
\bibitem[{\citenamefont{Lee et~al.}(2011{\natexlab{a}})\citenamefont{Lee, Raby,
  Ratz, Ross, Schieren et~al.}}]{ZNR1}
\bibinfo{author}{\bibfnamefont{H.~M.} \bibnamefont{Lee}},
  \bibinfo{author}{\bibfnamefont{S.}~\bibnamefont{Raby}},
  \bibinfo{author}{\bibfnamefont{M.}~\bibnamefont{Ratz}},
  \bibinfo{author}{\bibfnamefont{G.~G.} \bibnamefont{Ross}},
  \bibinfo{author}{\bibfnamefont{R.}~\bibnamefont{Schieren}},
  \bibnamefont{et~al.}, \bibinfo{journal}{Phys.Lett.}
  \textbf{\bibinfo{volume}{B694}}, \bibinfo{pages}{491}
  (\bibinfo{year}{2011}{\natexlab{a}}), \eprint{1009.0905}\relax
\mciteBstWouldAddEndPuncttrue
\mciteSetBstMidEndSepPunct{\mcitedefaultmidpunct}
{\mcitedefaultendpunct}{\mcitedefaultseppunct}\relax
\EndOfBibitem
\bibitem[{\citenamefont{Lee et~al.}(2011{\natexlab{b}})\citenamefont{Lee, Raby,
  Ratz, Ross, Schieren et~al.}}]{ZNR2}
\bibinfo{author}{\bibfnamefont{H.~M.} \bibnamefont{Lee}},
  \bibinfo{author}{\bibfnamefont{S.}~\bibnamefont{Raby}},
  \bibinfo{author}{\bibfnamefont{M.}~\bibnamefont{Ratz}},
  \bibinfo{author}{\bibfnamefont{G.~G.} \bibnamefont{Ross}},
  \bibinfo{author}{\bibfnamefont{R.}~\bibnamefont{Schieren}},
  \bibnamefont{et~al.}, \bibinfo{journal}{Nucl.Phys.}
  \textbf{\bibinfo{volume}{B850}}, \bibinfo{pages}{1}
  (\bibinfo{year}{2011}{\natexlab{b}}), \eprint{1102.3595}\relax
\mciteBstWouldAddEndPuncttrue
\mciteSetBstMidEndSepPunct{\mcitedefaultmidpunct}
{\mcitedefaultendpunct}{\mcitedefaultseppunct}\relax
\EndOfBibitem
\bibitem[{\citenamefont{Froggatt and Nielsen}(1979)}]{Froggatt:1978nt}
\bibinfo{author}{\bibfnamefont{C.}~\bibnamefont{Froggatt}} \bibnamefont{and}
  \bibinfo{author}{\bibfnamefont{H.~B.} \bibnamefont{Nielsen}},
  \bibinfo{journal}{Nucl.Phys.} \textbf{\bibinfo{volume}{B147}},
  \bibinfo{pages}{277} (\bibinfo{year}{1979})\relax
\mciteBstWouldAddEndPuncttrue
\mciteSetBstMidEndSepPunct{\mcitedefaultmidpunct}
{\mcitedefaultendpunct}{\mcitedefaultseppunct}\relax
\EndOfBibitem
\bibitem[{\citenamefont{King and Luhn}(2009)}]{KingPSL1}
\bibinfo{author}{\bibfnamefont{S.~F.} \bibnamefont{King}} \bibnamefont{and}
  \bibinfo{author}{\bibfnamefont{C.}~\bibnamefont{Luhn}},
  \bibinfo{journal}{Nucl.Phys.} \textbf{\bibinfo{volume}{B820}},
  \bibinfo{pages}{269} (\bibinfo{year}{2009}), \eprint{0905.1686}\relax
\mciteBstWouldAddEndPuncttrue
\mciteSetBstMidEndSepPunct{\mcitedefaultmidpunct}
{\mcitedefaultendpunct}{\mcitedefaultseppunct}\relax
\EndOfBibitem
\bibitem[{\citenamefont{King and Luhn}(2010)}]{KingPSL2}
\bibinfo{author}{\bibfnamefont{S.~F.} \bibnamefont{King}} \bibnamefont{and}
  \bibinfo{author}{\bibfnamefont{C.}~\bibnamefont{Luhn}},
  \bibinfo{journal}{Nucl.Phys.} \textbf{\bibinfo{volume}{B832}},
  \bibinfo{pages}{414} (\bibinfo{year}{2010}), \eprint{0912.1344}\relax
\mciteBstWouldAddEndPuncttrue
\mciteSetBstMidEndSepPunct{\mcitedefaultmidpunct}
{\mcitedefaultendpunct}{\mcitedefaultseppunct}\relax
\EndOfBibitem
\bibitem[{\citenamefont{de~Medeiros~Varzielas}(2012)}]{SU3SU3}
\bibinfo{author}{\bibfnamefont{I.}~\bibnamefont{de~Medeiros~Varzielas}},
  \bibinfo{journal}{JHEP} \textbf{\bibinfo{volume}{1201}}, \bibinfo{pages}{097}
  (\bibinfo{year}{2012}), \eprint{1111.3952}\relax
\mciteBstWouldAddEndPuncttrue
\mciteSetBstMidEndSepPunct{\mcitedefaultmidpunct}
{\mcitedefaultendpunct}{\mcitedefaultseppunct}\relax
\EndOfBibitem
\bibitem[{\citenamefont{Roberts et~al.}(2001)\citenamefont{Roberts, Romanino,
  Ross, and Velasco-Sevilla}}]{RRRV}
\bibinfo{author}{\bibfnamefont{R.}~\bibnamefont{Roberts}},
  \bibinfo{author}{\bibfnamefont{A.}~\bibnamefont{Romanino}},
  \bibinfo{author}{\bibfnamefont{G.~G.} \bibnamefont{Ross}}, \bibnamefont{and}
  \bibinfo{author}{\bibfnamefont{L.}~\bibnamefont{Velasco-Sevilla}},
  \bibinfo{journal}{Nucl.Phys.} \textbf{\bibinfo{volume}{B615}},
  \bibinfo{pages}{358} (\bibinfo{year}{2001}), \eprint{hep-ph/0104088}\relax
\mciteBstWouldAddEndPuncttrue
\mciteSetBstMidEndSepPunct{\mcitedefaultmidpunct}
{\mcitedefaultendpunct}{\mcitedefaultseppunct}\relax
\EndOfBibitem
\bibitem[{\citenamefont{Ross and Serna}(2008)}]{Mario}
\bibinfo{author}{\bibfnamefont{G.}~\bibnamefont{Ross}} \bibnamefont{and}
  \bibinfo{author}{\bibfnamefont{M.}~\bibnamefont{Serna}},
  \bibinfo{journal}{Phys.Lett.} \textbf{\bibinfo{volume}{B664}},
  \bibinfo{pages}{97} (\bibinfo{year}{2008}), \eprint{0704.1248}\relax
\mciteBstWouldAddEndPuncttrue
\mciteSetBstMidEndSepPunct{\mcitedefaultmidpunct}
{\mcitedefaultendpunct}{\mcitedefaultseppunct}\relax
\EndOfBibitem
\bibitem[{\citenamefont{King}(2000)}]{King:1999mb}
\bibinfo{author}{\bibfnamefont{S.}~\bibnamefont{King}},
  \bibinfo{journal}{Nucl.Phys.} \textbf{\bibinfo{volume}{B576}},
  \bibinfo{pages}{85} (\bibinfo{year}{2000}), \eprint{hep-ph/9912492}\relax
\mciteBstWouldAddEndPuncttrue
\mciteSetBstMidEndSepPunct{\mcitedefaultmidpunct}
{\mcitedefaultendpunct}{\mcitedefaultseppunct}\relax
\EndOfBibitem
\bibitem[{\citenamefont{Abe et~al.}(2011)}]{Abe:2011sj}
\bibinfo{author}{\bibfnamefont{K.}~\bibnamefont{Abe}} \bibnamefont{et~al.}
  (\bibinfo{collaboration}{T2K}), \bibinfo{journal}{Phys. Rev. Lett.}
  \textbf{\bibinfo{volume}{107}}, \bibinfo{pages}{041801}
  (\bibinfo{year}{2011}), \eprint{1106.2822}\relax
\mciteBstWouldAddEndPuncttrue
\mciteSetBstMidEndSepPunct{\mcitedefaultmidpunct}
{\mcitedefaultendpunct}{\mcitedefaultseppunct}\relax
\EndOfBibitem
\bibitem[{\citenamefont{Fogli et~al.}(2011)\citenamefont{Fogli, Lisi, Marrone,
  Palazzo, and Rotunno}}]{Fogli:2011qn}
\bibinfo{author}{\bibfnamefont{G.~L.} \bibnamefont{Fogli}},
  \bibinfo{author}{\bibfnamefont{E.}~\bibnamefont{Lisi}},
  \bibinfo{author}{\bibfnamefont{A.}~\bibnamefont{Marrone}},
  \bibinfo{author}{\bibfnamefont{A.}~\bibnamefont{Palazzo}}, \bibnamefont{and}
  \bibinfo{author}{\bibfnamefont{A.~M.} \bibnamefont{Rotunno}},
  \bibinfo{journal}{Phys. Rev.} \textbf{\bibinfo{volume}{D84}},
  \bibinfo{pages}{053007} (\bibinfo{year}{2011}), \eprint{1106.6028}\relax
\mciteBstWouldAddEndPuncttrue
\mciteSetBstMidEndSepPunct{\mcitedefaultmidpunct}
{\mcitedefaultendpunct}{\mcitedefaultseppunct}\relax
\EndOfBibitem
\bibitem[{\citenamefont{Schwetz et~al.}(2011)\citenamefont{Schwetz, Tortola,
  and Valle}}]{Schwetz:2011zk}
\bibinfo{author}{\bibfnamefont{T.}~\bibnamefont{Schwetz}},
  \bibinfo{author}{\bibfnamefont{M.}~\bibnamefont{Tortola}}, \bibnamefont{and}
  \bibinfo{author}{\bibfnamefont{J.~W.~F.} \bibnamefont{Valle}},
  \bibinfo{journal}{New J. Phys.} \textbf{\bibinfo{volume}{13}},
  \bibinfo{pages}{109401} (\bibinfo{year}{2011}), \eprint{1108.1376}\relax
\mciteBstWouldAddEndPuncttrue
\mciteSetBstMidEndSepPunct{\mcitedefaultmidpunct}
{\mcitedefaultendpunct}{\mcitedefaultseppunct}\relax
\EndOfBibitem
\bibitem[{\citenamefont{Machado et~al.}(2011)\citenamefont{Machado, Minakata,
  Nunokawa, and Funchal}}]{Machado:2011ar}
\bibinfo{author}{\bibfnamefont{P.}~\bibnamefont{Machado}},
  \bibinfo{author}{\bibfnamefont{H.}~\bibnamefont{Minakata}},
  \bibinfo{author}{\bibfnamefont{H.}~\bibnamefont{Nunokawa}}, \bibnamefont{and}
  \bibinfo{author}{\bibfnamefont{R.~Z.} \bibnamefont{Funchal}}
  (\bibinfo{year}{2011}), \eprint{1111.3330}\relax
\mciteBstWouldAddEndPuncttrue
\mciteSetBstMidEndSepPunct{\mcitedefaultmidpunct}
{\mcitedefaultendpunct}{\mcitedefaultseppunct}\relax
\EndOfBibitem
\bibitem[{\citenamefont{An et~al.}(2012)}]{DayaBay}
\bibinfo{author}{\bibfnamefont{F.}~\bibnamefont{An}} \bibnamefont{et~al.}
  (\bibinfo{collaboration}{DAYA-BAY Collaboration}) (\bibinfo{year}{2012}),
  \eprint{1203.1669}\relax
\mciteBstWouldAddEndPuncttrue
\mciteSetBstMidEndSepPunct{\mcitedefaultmidpunct}
{\mcitedefaultendpunct}{\mcitedefaultseppunct}\relax
\EndOfBibitem
\bibitem[{\citenamefont{He and Yin}(2011)}]{He:2011kn}
\bibinfo{author}{\bibfnamefont{H.-J.} \bibnamefont{He}} \bibnamefont{and}
  \bibinfo{author}{\bibfnamefont{F.-R.} \bibnamefont{Yin}},
  \bibinfo{journal}{Phys.Rev.} \textbf{\bibinfo{volume}{D84}},
  \bibinfo{pages}{033009} (\bibinfo{year}{2011}), \eprint{1104.2654}\relax
\mciteBstWouldAddEndPuncttrue
\mciteSetBstMidEndSepPunct{\mcitedefaultmidpunct}
{\mcitedefaultendpunct}{\mcitedefaultseppunct}\relax
\EndOfBibitem
\bibitem[{\citenamefont{Xing}(2011)}]{Xing:2011at}
\bibinfo{author}{\bibfnamefont{Z.-z.} \bibnamefont{Xing}}
  (\bibinfo{year}{2011}), \eprint{1106.3244}\relax
\mciteBstWouldAddEndPuncttrue
\mciteSetBstMidEndSepPunct{\mcitedefaultmidpunct}
{\mcitedefaultendpunct}{\mcitedefaultseppunct}\relax
\EndOfBibitem
\bibitem[{\citenamefont{Ma and Wegman}(2011)}]{Ma:2011yi}
\bibinfo{author}{\bibfnamefont{E.}~\bibnamefont{Ma}} \bibnamefont{and}
  \bibinfo{author}{\bibfnamefont{D.}~\bibnamefont{Wegman}},
  \bibinfo{journal}{Phys.Rev.Lett.} \textbf{\bibinfo{volume}{107}},
  \bibinfo{pages}{061803} (\bibinfo{year}{2011}), \eprint{1106.4269}\relax
\mciteBstWouldAddEndPuncttrue
\mciteSetBstMidEndSepPunct{\mcitedefaultmidpunct}
{\mcitedefaultendpunct}{\mcitedefaultseppunct}\relax
\EndOfBibitem
\bibitem[{\citenamefont{Zheng and Ma}(2012)}]{Zheng:2011uz}
\bibinfo{author}{\bibfnamefont{Y.-j.} \bibnamefont{Zheng}} \bibnamefont{and}
  \bibinfo{author}{\bibfnamefont{B.-Q.} \bibnamefont{Ma}},
  \bibinfo{journal}{Eur. Phys. J.} \textbf{\bibinfo{volume}{Plus}},
  \bibinfo{pages}{127(1): 7} (\bibinfo{year}{2012}), \eprint{1106.4040}\relax
\mciteBstWouldAddEndPuncttrue
\mciteSetBstMidEndSepPunct{\mcitedefaultmidpunct}
{\mcitedefaultendpunct}{\mcitedefaultseppunct}\relax
\EndOfBibitem
\bibitem[{\citenamefont{Zhou}(2011)}]{Zhou:2011nu}
\bibinfo{author}{\bibfnamefont{S.}~\bibnamefont{Zhou}},
  \bibinfo{journal}{Phys.Lett.} \textbf{\bibinfo{volume}{B704}},
  \bibinfo{pages}{291} (\bibinfo{year}{2011}), \eprint{1106.4808}\relax
\mciteBstWouldAddEndPuncttrue
\mciteSetBstMidEndSepPunct{\mcitedefaultmidpunct}
{\mcitedefaultendpunct}{\mcitedefaultseppunct}\relax
\EndOfBibitem
\bibitem[{\citenamefont{Araki}(2011)}]{Araki:2011wn}
\bibinfo{author}{\bibfnamefont{T.}~\bibnamefont{Araki}},
  \bibinfo{journal}{Phys.Rev.} \textbf{\bibinfo{volume}{D84}},
  \bibinfo{pages}{037301} (\bibinfo{year}{2011}), \eprint{1106.5211}\relax
\mciteBstWouldAddEndPuncttrue
\mciteSetBstMidEndSepPunct{\mcitedefaultmidpunct}
{\mcitedefaultendpunct}{\mcitedefaultseppunct}\relax
\EndOfBibitem
\bibitem[{\citenamefont{Haba and Takahashi}(2011)}]{Haba:2011nv}
\bibinfo{author}{\bibfnamefont{N.}~\bibnamefont{Haba}} \bibnamefont{and}
  \bibinfo{author}{\bibfnamefont{R.}~\bibnamefont{Takahashi}},
  \bibinfo{journal}{Phys.Lett.} \textbf{\bibinfo{volume}{B702}},
  \bibinfo{pages}{388} (\bibinfo{year}{2011}), \eprint{1106.5926}\relax
\mciteBstWouldAddEndPuncttrue
\mciteSetBstMidEndSepPunct{\mcitedefaultmidpunct}
{\mcitedefaultendpunct}{\mcitedefaultseppunct}\relax
\EndOfBibitem
\bibitem[{\citenamefont{Morisi et~al.}(2011)\citenamefont{Morisi, Patel, and
  Peinado}}]{Morisi:2011pm}
\bibinfo{author}{\bibfnamefont{S.}~\bibnamefont{Morisi}},
  \bibinfo{author}{\bibfnamefont{K.~M.} \bibnamefont{Patel}}, \bibnamefont{and}
  \bibinfo{author}{\bibfnamefont{E.}~\bibnamefont{Peinado}},
  \bibinfo{journal}{Phys.Rev.} \textbf{\bibinfo{volume}{D84}},
  \bibinfo{pages}{053002} (\bibinfo{year}{2011}), \eprint{1107.0696}\relax
\mciteBstWouldAddEndPuncttrue
\mciteSetBstMidEndSepPunct{\mcitedefaultmidpunct}
{\mcitedefaultendpunct}{\mcitedefaultseppunct}\relax
\EndOfBibitem
\bibitem[{\citenamefont{Chao and Zheng}(2011)}]{Chao:2011sp}
\bibinfo{author}{\bibfnamefont{W.}~\bibnamefont{Chao}} \bibnamefont{and}
  \bibinfo{author}{\bibfnamefont{Y.-j.} \bibnamefont{Zheng}}
  (\bibinfo{year}{2011}), \eprint{1107.0738}\relax
\mciteBstWouldAddEndPuncttrue
\mciteSetBstMidEndSepPunct{\mcitedefaultmidpunct}
{\mcitedefaultendpunct}{\mcitedefaultseppunct}\relax
\EndOfBibitem
\bibitem[{\citenamefont{Zhang and Zhou}(2011)}]{Zhang:2011aw}
\bibinfo{author}{\bibfnamefont{H.}~\bibnamefont{Zhang}} \bibnamefont{and}
  \bibinfo{author}{\bibfnamefont{S.}~\bibnamefont{Zhou}},
  \bibinfo{journal}{Phys.Lett.} \textbf{\bibinfo{volume}{B704}},
  \bibinfo{pages}{296} (\bibinfo{year}{2011}), \eprint{1107.1097}\relax
\mciteBstWouldAddEndPuncttrue
\mciteSetBstMidEndSepPunct{\mcitedefaultmidpunct}
{\mcitedefaultendpunct}{\mcitedefaultseppunct}\relax
\EndOfBibitem
\bibitem[{\citenamefont{Dev et~al.}(2011{\natexlab{a}})\citenamefont{Dev,
  Gupta, and Gautam}}]{Dev:2011bd}
\bibinfo{author}{\bibfnamefont{S.}~\bibnamefont{Dev}},
  \bibinfo{author}{\bibfnamefont{S.}~\bibnamefont{Gupta}}, \bibnamefont{and}
  \bibinfo{author}{\bibfnamefont{R.~R.} \bibnamefont{Gautam}},
  \bibinfo{journal}{Phys.Lett.} \textbf{\bibinfo{volume}{B704}},
  \bibinfo{pages}{527} (\bibinfo{year}{2011}{\natexlab{a}}),
  \eprint{1107.1125}\relax
\mciteBstWouldAddEndPuncttrue
\mciteSetBstMidEndSepPunct{\mcitedefaultmidpunct}
{\mcitedefaultendpunct}{\mcitedefaultseppunct}\relax
\EndOfBibitem
\bibitem[{\citenamefont{Chu et~al.}(2011)\citenamefont{Chu, Dhen, and
  Hambye}}]{Chu:2011jg}
\bibinfo{author}{\bibfnamefont{X.}~\bibnamefont{Chu}},
  \bibinfo{author}{\bibfnamefont{M.}~\bibnamefont{Dhen}}, \bibnamefont{and}
  \bibinfo{author}{\bibfnamefont{T.}~\bibnamefont{Hambye}},
  \bibinfo{journal}{JHEP} \textbf{\bibinfo{volume}{1111}}, \bibinfo{pages}{106}
  (\bibinfo{year}{2011}), \eprint{1107.1589}\relax
\mciteBstWouldAddEndPuncttrue
\mciteSetBstMidEndSepPunct{\mcitedefaultmidpunct}
{\mcitedefaultendpunct}{\mcitedefaultseppunct}\relax
\EndOfBibitem
\bibitem[{\citenamefont{Bhupal~Dev et~al.}(2011)\citenamefont{Bhupal~Dev,
  Mohapatra, and Severson}}]{BhupalDev:2011gi}
\bibinfo{author}{\bibfnamefont{P.}~\bibnamefont{Bhupal~Dev}},
  \bibinfo{author}{\bibfnamefont{R.}~\bibnamefont{Mohapatra}},
  \bibnamefont{and} \bibinfo{author}{\bibfnamefont{M.}~\bibnamefont{Severson}},
  \bibinfo{journal}{Phys.Rev.} \textbf{\bibinfo{volume}{D84}},
  \bibinfo{pages}{053005} (\bibinfo{year}{2011}), \eprint{1107.2378}\relax
\mciteBstWouldAddEndPuncttrue
\mciteSetBstMidEndSepPunct{\mcitedefaultmidpunct}
{\mcitedefaultendpunct}{\mcitedefaultseppunct}\relax
\EndOfBibitem
\bibitem[{\citenamefont{Toorop et~al.}(2011)\citenamefont{Toorop, Feruglio, and
  Hagedorn}}]{Toorop:2011jn}
\bibinfo{author}{\bibfnamefont{R.~d.~A.} \bibnamefont{Toorop}},
  \bibinfo{author}{\bibfnamefont{F.}~\bibnamefont{Feruglio}}, \bibnamefont{and}
  \bibinfo{author}{\bibfnamefont{C.}~\bibnamefont{Hagedorn}},
  \bibinfo{journal}{Phys.Lett.} \textbf{\bibinfo{volume}{B703}},
  \bibinfo{pages}{447} (\bibinfo{year}{2011}), \eprint{1107.3486}\relax
\mciteBstWouldAddEndPuncttrue
\mciteSetBstMidEndSepPunct{\mcitedefaultmidpunct}
{\mcitedefaultendpunct}{\mcitedefaultseppunct}\relax
\EndOfBibitem
\bibitem[{\citenamefont{Antusch and Maurer}(2011)}]{Antusch:2011qg}
\bibinfo{author}{\bibfnamefont{S.}~\bibnamefont{Antusch}} \bibnamefont{and}
  \bibinfo{author}{\bibfnamefont{V.}~\bibnamefont{Maurer}},
  \bibinfo{journal}{Phys.Rev.} \textbf{\bibinfo{volume}{D84}},
  \bibinfo{pages}{117301} (\bibinfo{year}{2011}), \eprint{1107.3728}\relax
\mciteBstWouldAddEndPuncttrue
\mciteSetBstMidEndSepPunct{\mcitedefaultmidpunct}
{\mcitedefaultendpunct}{\mcitedefaultseppunct}\relax
\EndOfBibitem
\bibitem[{\citenamefont{Rodejohann et~al.}(2012)\citenamefont{Rodejohann,
  Zhang, and Zhou}}]{Rodejohann:2011uz}
\bibinfo{author}{\bibfnamefont{W.}~\bibnamefont{Rodejohann}},
  \bibinfo{author}{\bibfnamefont{H.}~\bibnamefont{Zhang}}, \bibnamefont{and}
  \bibinfo{author}{\bibfnamefont{S.}~\bibnamefont{Zhou}},
  \bibinfo{journal}{Nucl.Phys.} \textbf{\bibinfo{volume}{B855}},
  \bibinfo{pages}{592} (\bibinfo{year}{2012}), \eprint{1107.3970}\relax
\mciteBstWouldAddEndPuncttrue
\mciteSetBstMidEndSepPunct{\mcitedefaultmidpunct}
{\mcitedefaultendpunct}{\mcitedefaultseppunct}\relax
\EndOfBibitem
\bibitem[{\citenamefont{Ahn et~al.}(2011)\citenamefont{Ahn, Cheng, and
  Oh}}]{Ahn:2011if}
\bibinfo{author}{\bibfnamefont{Y.}~\bibnamefont{Ahn}},
  \bibinfo{author}{\bibfnamefont{H.-Y.} \bibnamefont{Cheng}}, \bibnamefont{and}
  \bibinfo{author}{\bibfnamefont{S.}~\bibnamefont{Oh}},
  \bibinfo{journal}{Phys.Rev.} \textbf{\bibinfo{volume}{D84}},
  \bibinfo{pages}{113007} (\bibinfo{year}{2011}), \eprint{1107.4549}\relax
\mciteBstWouldAddEndPuncttrue
\mciteSetBstMidEndSepPunct{\mcitedefaultmidpunct}
{\mcitedefaultendpunct}{\mcitedefaultseppunct}\relax
\EndOfBibitem
\bibitem[{\citenamefont{King and Luhn}(2011{\natexlab{a}})}]{King:2011zj}
\bibinfo{author}{\bibfnamefont{S.~F.} \bibnamefont{King}} \bibnamefont{and}
  \bibinfo{author}{\bibfnamefont{C.}~\bibnamefont{Luhn}},
  \bibinfo{journal}{JHEP} \textbf{\bibinfo{volume}{1109}}, \bibinfo{pages}{042}
  (\bibinfo{year}{2011}{\natexlab{a}}), \eprint{1107.5332}\relax
\mciteBstWouldAddEndPuncttrue
\mciteSetBstMidEndSepPunct{\mcitedefaultmidpunct}
{\mcitedefaultendpunct}{\mcitedefaultseppunct}\relax
\EndOfBibitem
\bibitem[{\citenamefont{Marzocca et~al.}(2011)\citenamefont{Marzocca, Petcov,
  Romanino, and Spinrath}}]{Marzocca:2011dh}
\bibinfo{author}{\bibfnamefont{D.}~\bibnamefont{Marzocca}},
  \bibinfo{author}{\bibfnamefont{S.~T.} \bibnamefont{Petcov}},
  \bibinfo{author}{\bibfnamefont{A.}~\bibnamefont{Romanino}}, \bibnamefont{and}
  \bibinfo{author}{\bibfnamefont{M.}~\bibnamefont{Spinrath}},
  \bibinfo{journal}{JHEP} \textbf{\bibinfo{volume}{1111}}, \bibinfo{pages}{009}
  (\bibinfo{year}{2011}), \eprint{1108.0614}\relax
\mciteBstWouldAddEndPuncttrue
\mciteSetBstMidEndSepPunct{\mcitedefaultmidpunct}
{\mcitedefaultendpunct}{\mcitedefaultseppunct}\relax
\EndOfBibitem
\bibitem[{\citenamefont{Ge et~al.}(2011)\citenamefont{Ge, Dicus, and
  Repko}}]{Ge:2011qn}
\bibinfo{author}{\bibfnamefont{S.-F.} \bibnamefont{Ge}},
  \bibinfo{author}{\bibfnamefont{D.~A.} \bibnamefont{Dicus}}, \bibnamefont{and}
  \bibinfo{author}{\bibfnamefont{W.~W.} \bibnamefont{Repko}}
  (\bibinfo{year}{2011}), \eprint{1108.0964}\relax
\mciteBstWouldAddEndPuncttrue
\mciteSetBstMidEndSepPunct{\mcitedefaultmidpunct}
{\mcitedefaultendpunct}{\mcitedefaultseppunct}\relax
\EndOfBibitem
\bibitem[{\citenamefont{Kumar}(2011)}]{Kumar:2011vf}
\bibinfo{author}{\bibfnamefont{S.}~\bibnamefont{Kumar}},
  \bibinfo{journal}{Phys.Rev.} \textbf{\bibinfo{volume}{D84}},
  \bibinfo{pages}{077301} (\bibinfo{year}{2011}), \eprint{1108.2137}\relax
\mciteBstWouldAddEndPuncttrue
\mciteSetBstMidEndSepPunct{\mcitedefaultmidpunct}
{\mcitedefaultendpunct}{\mcitedefaultseppunct}\relax
\EndOfBibitem
\bibitem[{\citenamefont{Bazzocchi}(2011)}]{Bazzocchi:2011ax}
\bibinfo{author}{\bibfnamefont{F.}~\bibnamefont{Bazzocchi}}
  (\bibinfo{year}{2011}), \eprint{1108.2497}\relax
\mciteBstWouldAddEndPuncttrue
\mciteSetBstMidEndSepPunct{\mcitedefaultmidpunct}
{\mcitedefaultendpunct}{\mcitedefaultseppunct}\relax
\EndOfBibitem
\bibitem[{\citenamefont{Araki and Geng}(2011)}]{Araki:2011qy}
\bibinfo{author}{\bibfnamefont{T.}~\bibnamefont{Araki}} \bibnamefont{and}
  \bibinfo{author}{\bibfnamefont{C.-Q.} \bibnamefont{Geng}},
  \bibinfo{journal}{JHEP} \textbf{\bibinfo{volume}{1109}}, \bibinfo{pages}{139}
  (\bibinfo{year}{2011}), \eprint{1108.3175}\relax
\mciteBstWouldAddEndPuncttrue
\mciteSetBstMidEndSepPunct{\mcitedefaultmidpunct}
{\mcitedefaultendpunct}{\mcitedefaultseppunct}\relax
\EndOfBibitem
\bibitem[{\citenamefont{Antusch et~al.}(2012)\citenamefont{Antusch, King, Luhn,
  and Spinrath}}]{Antusch:2011ic}
\bibinfo{author}{\bibfnamefont{S.}~\bibnamefont{Antusch}},
  \bibinfo{author}{\bibfnamefont{S.~F.} \bibnamefont{King}},
  \bibinfo{author}{\bibfnamefont{C.}~\bibnamefont{Luhn}}, \bibnamefont{and}
  \bibinfo{author}{\bibfnamefont{M.}~\bibnamefont{Spinrath}},
  \bibinfo{journal}{Nucl.Phys.} \textbf{\bibinfo{volume}{B856}},
  \bibinfo{pages}{328} (\bibinfo{year}{2012}), \eprint{1108.4278}\relax
\mciteBstWouldAddEndPuncttrue
\mciteSetBstMidEndSepPunct{\mcitedefaultmidpunct}
{\mcitedefaultendpunct}{\mcitedefaultseppunct}\relax
\EndOfBibitem
\bibitem[{\citenamefont{Fritzsch et~al.}(2011)\citenamefont{Fritzsch, Xing, and
  Zhou}}]{Fritzsch:2011qv}
\bibinfo{author}{\bibfnamefont{H.}~\bibnamefont{Fritzsch}},
  \bibinfo{author}{\bibfnamefont{Z.-z.} \bibnamefont{Xing}}, \bibnamefont{and}
  \bibinfo{author}{\bibfnamefont{S.}~\bibnamefont{Zhou}},
  \bibinfo{journal}{JHEP} \textbf{\bibinfo{volume}{1109}}, \bibinfo{pages}{083}
  (\bibinfo{year}{2011}), \eprint{1108.4534}\relax
\mciteBstWouldAddEndPuncttrue
\mciteSetBstMidEndSepPunct{\mcitedefaultmidpunct}
{\mcitedefaultendpunct}{\mcitedefaultseppunct}\relax
\EndOfBibitem
\bibitem[{\citenamefont{Rashed and Datta}(2011)}]{Rashed:2011zs}
\bibinfo{author}{\bibfnamefont{A.}~\bibnamefont{Rashed}} \bibnamefont{and}
  \bibinfo{author}{\bibfnamefont{A.}~\bibnamefont{Datta}}
  (\bibinfo{year}{2011}), \eprint{1109.2320}\relax
\mciteBstWouldAddEndPuncttrue
\mciteSetBstMidEndSepPunct{\mcitedefaultmidpunct}
{\mcitedefaultendpunct}{\mcitedefaultseppunct}\relax
\EndOfBibitem
\bibitem[{\citenamefont{Ludl et~al.}(2012)\citenamefont{Ludl, Morisi, and
  Peinado}}]{Ludl:2011vv}
\bibinfo{author}{\bibfnamefont{P.}~\bibnamefont{Ludl}},
  \bibinfo{author}{\bibfnamefont{S.}~\bibnamefont{Morisi}}, \bibnamefont{and}
  \bibinfo{author}{\bibfnamefont{E.}~\bibnamefont{Peinado}},
  \bibinfo{journal}{Nucl.Phys.} \textbf{\bibinfo{volume}{B857}},
  \bibinfo{pages}{411} (\bibinfo{year}{2012}), \eprint{1109.3393}\relax
\mciteBstWouldAddEndPuncttrue
\mciteSetBstMidEndSepPunct{\mcitedefaultmidpunct}
{\mcitedefaultendpunct}{\mcitedefaultseppunct}\relax
\EndOfBibitem
\bibitem[{\citenamefont{Verma}(2012)}]{Verma:2011kz}
\bibinfo{author}{\bibfnamefont{S.}~\bibnamefont{Verma}},
  \bibinfo{journal}{Nucl.Phys.} \textbf{\bibinfo{volume}{B854}},
  \bibinfo{pages}{340} (\bibinfo{year}{2012}), \eprint{1109.4228}\relax
\mciteBstWouldAddEndPuncttrue
\mciteSetBstMidEndSepPunct{\mcitedefaultmidpunct}
{\mcitedefaultendpunct}{\mcitedefaultseppunct}\relax
\EndOfBibitem
\bibitem[{\citenamefont{Meloni}(2011)}]{Meloni:2011ac}
\bibinfo{author}{\bibfnamefont{D.}~\bibnamefont{Meloni}}
  (\bibinfo{year}{2011}), \eprint{1110.5210}\relax
\mciteBstWouldAddEndPuncttrue
\mciteSetBstMidEndSepPunct{\mcitedefaultmidpunct}
{\mcitedefaultendpunct}{\mcitedefaultseppunct}\relax
\EndOfBibitem
\bibitem[{\citenamefont{Dev et~al.}(2011{\natexlab{b}})\citenamefont{Dev,
  Gupta, Gautam, and Singh}}]{Dev:2011hf}
\bibinfo{author}{\bibfnamefont{S.}~\bibnamefont{Dev}},
  \bibinfo{author}{\bibfnamefont{S.}~\bibnamefont{Gupta}},
  \bibinfo{author}{\bibfnamefont{R.~R.} \bibnamefont{Gautam}},
  \bibnamefont{and} \bibinfo{author}{\bibfnamefont{L.}~\bibnamefont{Singh}},
  \bibinfo{journal}{Phys.Lett.} \textbf{\bibinfo{volume}{B706}},
  \bibinfo{pages}{168} (\bibinfo{year}{2011}{\natexlab{b}}),
  \eprint{1111.1300}\relax
\mciteBstWouldAddEndPuncttrue
\mciteSetBstMidEndSepPunct{\mcitedefaultmidpunct}
{\mcitedefaultendpunct}{\mcitedefaultseppunct}\relax
\EndOfBibitem
\bibitem[{\citenamefont{Deepthi et~al.}(2011)\citenamefont{Deepthi, Gollu, and
  Mohanta}}]{Deepthi:2011sk}
\bibinfo{author}{\bibfnamefont{K.}~\bibnamefont{Deepthi}},
  \bibinfo{author}{\bibfnamefont{S.}~\bibnamefont{Gollu}}, \bibnamefont{and}
  \bibinfo{author}{\bibfnamefont{R.}~\bibnamefont{Mohanta}}
  (\bibinfo{year}{2011}), \eprint{1111.2781}\relax
\mciteBstWouldAddEndPuncttrue
\mciteSetBstMidEndSepPunct{\mcitedefaultmidpunct}
{\mcitedefaultendpunct}{\mcitedefaultseppunct}\relax
\EndOfBibitem
\bibitem[{\citenamefont{Rashed}(2011)}]{Rashed:2011xe}
\bibinfo{author}{\bibfnamefont{A.}~\bibnamefont{Rashed}}
  (\bibinfo{year}{2011}), \eprint{1111.3072}\relax
\mciteBstWouldAddEndPuncttrue
\mciteSetBstMidEndSepPunct{\mcitedefaultmidpunct}
{\mcitedefaultendpunct}{\mcitedefaultseppunct}\relax
\EndOfBibitem
\bibitem[{\citenamefont{King and Luhn}(2011{\natexlab{b}})}]{King:2011ab}
\bibinfo{author}{\bibfnamefont{S.~F.} \bibnamefont{King}} \bibnamefont{and}
  \bibinfo{author}{\bibfnamefont{C.}~\bibnamefont{Luhn}}
  (\bibinfo{year}{2011}{\natexlab{b}}), \eprint{1112.1959}\relax
\mciteBstWouldAddEndPuncttrue
\mciteSetBstMidEndSepPunct{\mcitedefaultmidpunct}
{\mcitedefaultendpunct}{\mcitedefaultseppunct}\relax
\EndOfBibitem
\bibitem[{\citenamefont{Araki and Li}(2011)}]{Araki:2011zg}
\bibinfo{author}{\bibfnamefont{T.}~\bibnamefont{Araki}} \bibnamefont{and}
  \bibinfo{author}{\bibfnamefont{Y.}~\bibnamefont{Li}} (\bibinfo{year}{2011}),
  \eprint{1112.5819}\relax
\mciteBstWouldAddEndPuncttrue
\mciteSetBstMidEndSepPunct{\mcitedefaultmidpunct}
{\mcitedefaultendpunct}{\mcitedefaultseppunct}\relax
\EndOfBibitem
\bibitem[{\citenamefont{Gupta et~al.}(2011)\citenamefont{Gupta, Joshipura, and
  Patel}}]{Gupta:2011ct}
\bibinfo{author}{\bibfnamefont{S.}~\bibnamefont{Gupta}},
  \bibinfo{author}{\bibfnamefont{A.~S.} \bibnamefont{Joshipura}},
  \bibnamefont{and} \bibinfo{author}{\bibfnamefont{K.~M.} \bibnamefont{Patel}}
  (\bibinfo{year}{2011}), \eprint{1112.6113}\relax
\mciteBstWouldAddEndPuncttrue
\mciteSetBstMidEndSepPunct{\mcitedefaultmidpunct}
{\mcitedefaultendpunct}{\mcitedefaultseppunct}\relax
\EndOfBibitem
\bibitem[{\citenamefont{Ding}(2012)}]{Ding:2012xx}
\bibinfo{author}{\bibfnamefont{G.-J.} \bibnamefont{Ding}}
  (\bibinfo{year}{2012}), \eprint{1201.3279}\relax
\mciteBstWouldAddEndPuncttrue
\mciteSetBstMidEndSepPunct{\mcitedefaultmidpunct}
{\mcitedefaultendpunct}{\mcitedefaultseppunct}\relax
\EndOfBibitem
\bibitem[{\citenamefont{Ishimori and Kobayashi}(2012)}]{Ishimori:2012gv}
\bibinfo{author}{\bibfnamefont{H.}~\bibnamefont{Ishimori}} \bibnamefont{and}
  \bibinfo{author}{\bibfnamefont{T.}~\bibnamefont{Kobayashi}}
  (\bibinfo{year}{2012}), \eprint{1201.3429}\relax
\mciteBstWouldAddEndPuncttrue
\mciteSetBstMidEndSepPunct{\mcitedefaultmidpunct}
{\mcitedefaultendpunct}{\mcitedefaultseppunct}\relax
\EndOfBibitem
\bibitem[{\citenamefont{Dev et~al.}(2012)\citenamefont{Dev, Gautam, and
  Singh}}]{Dev:2012ns}
\bibinfo{author}{\bibfnamefont{S.}~\bibnamefont{Dev}},
  \bibinfo{author}{\bibfnamefont{R.~R.} \bibnamefont{Gautam}},
  \bibnamefont{and} \bibinfo{author}{\bibfnamefont{L.}~\bibnamefont{Singh}}
  (\bibinfo{year}{2012}), \eprint{1201.3755}\relax
\mciteBstWouldAddEndPuncttrue
\mciteSetBstMidEndSepPunct{\mcitedefaultmidpunct}
{\mcitedefaultendpunct}{\mcitedefaultseppunct}\relax
\EndOfBibitem
\bibitem[{\citenamefont{Bazzocchi et~al.}(2012)\citenamefont{Bazzocchi, Morisi,
  Peinado, Valle, and Vicente}}]{Bazzocchi:2012ve}
\bibinfo{author}{\bibfnamefont{F.}~\bibnamefont{Bazzocchi}},
  \bibinfo{author}{\bibfnamefont{S.}~\bibnamefont{Morisi}},
  \bibinfo{author}{\bibfnamefont{E.}~\bibnamefont{Peinado}},
  \bibinfo{author}{\bibfnamefont{J.}~\bibnamefont{Valle}}, \bibnamefont{and}
  \bibinfo{author}{\bibfnamefont{A.}~\bibnamefont{Vicente}}
  (\bibinfo{year}{2012}), \eprint{1202.1529}\relax
\mciteBstWouldAddEndPuncttrue
\mciteSetBstMidEndSepPunct{\mcitedefaultmidpunct}
{\mcitedefaultendpunct}{\mcitedefaultseppunct}\relax
\EndOfBibitem
\bibitem[{\citenamefont{Bhupal~Dev et~al.}(2012)\citenamefont{Bhupal~Dev,
  Dutta, Mohapatra, and Severson}}]{BhupalDev:2012nm}
\bibinfo{author}{\bibfnamefont{P.}~\bibnamefont{Bhupal~Dev}},
  \bibinfo{author}{\bibfnamefont{B.}~\bibnamefont{Dutta}},
  \bibinfo{author}{\bibfnamefont{R.}~\bibnamefont{Mohapatra}},
  \bibnamefont{and} \bibinfo{author}{\bibfnamefont{M.}~\bibnamefont{Severson}}
  (\bibinfo{year}{2012}), \eprint{1202.4012}\relax
\mciteBstWouldAddEndPuncttrue
\mciteSetBstMidEndSepPunct{\mcitedefaultmidpunct}
{\mcitedefaultendpunct}{\mcitedefaultseppunct}\relax
\EndOfBibitem
\bibitem[{\citenamefont{Cooper et~al.}(2012)\citenamefont{Cooper, King, and
  Luhn}}]{Cooper:2012wf}
\bibinfo{author}{\bibfnamefont{I.~K.} \bibnamefont{Cooper}},
  \bibinfo{author}{\bibfnamefont{S.~F.} \bibnamefont{King}}, \bibnamefont{and}
  \bibinfo{author}{\bibfnamefont{C.}~\bibnamefont{Luhn}}
  (\bibinfo{year}{2012}), \eprint{1203.1324}\relax
\mciteBstWouldAddEndPuncttrue
\mciteSetBstMidEndSepPunct{\mcitedefaultmidpunct}
{\mcitedefaultendpunct}{\mcitedefaultseppunct}\relax
\EndOfBibitem
\bibitem[{\citenamefont{Siyeon}(2012)}]{Siyeon:2012zu}
\bibinfo{author}{\bibfnamefont{K.}~\bibnamefont{Siyeon}}
  (\bibinfo{year}{2012}), \eprint{1203.1593}\relax
\mciteBstWouldAddEndPuncttrue
\mciteSetBstMidEndSepPunct{\mcitedefaultmidpunct}
{\mcitedefaultendpunct}{\mcitedefaultseppunct}\relax
\EndOfBibitem
\bibitem[{\citenamefont{Wu}(2012)}]{Wu:2012ri}
\bibinfo{author}{\bibfnamefont{Y.-L.} \bibnamefont{Wu}} (\bibinfo{year}{2012}),
  \eprint{1203.2382}\relax
\mciteBstWouldAddEndPuncttrue
\mciteSetBstMidEndSepPunct{\mcitedefaultmidpunct}
{\mcitedefaultendpunct}{\mcitedefaultseppunct}\relax
\EndOfBibitem
\bibitem[{\citenamefont{Branco et~al.}(2012)\citenamefont{Branco, Felipe,
  Joaquim, and Serodio}}]{Branco:2012vs}
\bibinfo{author}{\bibfnamefont{G.}~\bibnamefont{Branco}},
  \bibinfo{author}{\bibfnamefont{R.}~\bibnamefont{Felipe}},
  \bibinfo{author}{\bibfnamefont{F.}~\bibnamefont{Joaquim}}, \bibnamefont{and}
  \bibinfo{author}{\bibfnamefont{H.}~\bibnamefont{Serodio}}
  (\bibinfo{year}{2012}), \eprint{1203.2646}\relax
\mciteBstWouldAddEndPuncttrue
\mciteSetBstMidEndSepPunct{\mcitedefaultmidpunct}
{\mcitedefaultendpunct}{\mcitedefaultseppunct}\relax
\EndOfBibitem
\bibitem[{\citenamefont{He and Xu}(2012)}]{He:2012yt}
\bibinfo{author}{\bibfnamefont{H.-J.} \bibnamefont{He}} \bibnamefont{and}
  \bibinfo{author}{\bibfnamefont{X.-J.} \bibnamefont{Xu}}
  (\bibinfo{year}{2012}), \eprint{1203.2908}\relax
\mciteBstWouldAddEndPuncttrue
\mciteSetBstMidEndSepPunct{\mcitedefaultmidpunct}
{\mcitedefaultendpunct}{\mcitedefaultseppunct}\relax
\EndOfBibitem
\bibitem[{\citenamefont{Meloni}(2012)}]{Meloni:2012ci}
\bibinfo{author}{\bibfnamefont{D.}~\bibnamefont{Meloni}}
  (\bibinfo{year}{2012}), \eprint{1203.3126}\relax
\mciteBstWouldAddEndPuncttrue
\mciteSetBstMidEndSepPunct{\mcitedefaultmidpunct}
{\mcitedefaultendpunct}{\mcitedefaultseppunct}\relax
\EndOfBibitem
\bibitem[{\citenamefont{Ahn and Kang}(2012)}]{Ahn:2012tv}
\bibinfo{author}{\bibfnamefont{Y.}~\bibnamefont{Ahn}} \bibnamefont{and}
  \bibinfo{author}{\bibfnamefont{S.~K.} \bibnamefont{Kang}}
  (\bibinfo{year}{2012}), \eprint{1203.4185}\relax
\mciteBstWouldAddEndPuncttrue
\mciteSetBstMidEndSepPunct{\mcitedefaultmidpunct}
{\mcitedefaultendpunct}{\mcitedefaultseppunct}\relax
\EndOfBibitem
\bibitem[{\citenamefont{King}(2009)}]{King:2009qt}
\bibinfo{author}{\bibfnamefont{S.}~\bibnamefont{King}},
  \bibinfo{journal}{Phys.Lett.} \textbf{\bibinfo{volume}{B675}},
  \bibinfo{pages}{347} (\bibinfo{year}{2009}), \eprint{0903.3199}\relax
\mciteBstWouldAddEndPuncttrue
\mciteSetBstMidEndSepPunct{\mcitedefaultmidpunct}
{\mcitedefaultendpunct}{\mcitedefaultseppunct}\relax
\EndOfBibitem
\bibitem[{\citenamefont{Hall and Ross}(2012)}]{HallRoss}
\bibinfo{author}{\bibfnamefont{L.}~\bibnamefont{Hall}} \bibnamefont{and}
  \bibinfo{author}{\bibfnamefont{G.~G.} \bibnamefont{Ross}}
  (\bibinfo{year}{2012}), \bibinfo{note}{in preparation, presented at BeNe2012
  Trieste}\relax
\mciteBstWouldAddEndPuncttrue
\mciteSetBstMidEndSepPunct{\mcitedefaultmidpunct}
{\mcitedefaultendpunct}{\mcitedefaultseppunct}\relax
\EndOfBibitem
\bibitem[{\citenamefont{Ibanez and Ross}(1982)}]{IbanezRoss}
\bibinfo{author}{\bibfnamefont{L.~E.} \bibnamefont{Ibanez}} \bibnamefont{and}
  \bibinfo{author}{\bibfnamefont{G.~G.} \bibnamefont{Ross}},
  \bibinfo{journal}{Phys.Lett.} \textbf{\bibinfo{volume}{B110}},
  \bibinfo{pages}{215} (\bibinfo{year}{1982})\relax
\mciteBstWouldAddEndPuncttrue
\mciteSetBstMidEndSepPunct{\mcitedefaultmidpunct}
{\mcitedefaultendpunct}{\mcitedefaultseppunct}\relax
\EndOfBibitem
\end{mcitethebibliography}

\end{document}